\documentclass[lettersize,journal]{IEEEtran}
\usepackage{cite,color}
\usepackage{amsmath,amssymb,amsfonts}
\usepackage{algorithmic}
\usepackage{algorithm}
\usepackage{array}
\usepackage{url}
\usepackage[normalem]{ulem}
\useunder{\uline}{\ul}{}
\usepackage{float}
\usepackage{subfigure,multirow,bm,stfloats}
\usepackage{textcomp}
\usepackage{amsthm}
\newtheorem{definition}{Definition}

\newtheorem{property}{Property}
\usepackage{graphicx}

\begin{document}

\title{Signal Processing over {Multilayer Graphs}: Theoretical Foundations and Practical Applications}

\author{Songyang~Zhang, Qinwen~Deng,
	and~Zhi~Ding,~\IEEEmembership{Fellow,~IEEE}
\thanks{ This work was supported in part by the National Science Foundation under Grants 2029848 and 2002937.}
\thanks{S. Zhang, Q. Deng and Z. Ding are with Department of Electrical and Computer Engineering, University of California, Davis, CA, 95616. (E-mail: sydzhang@ucdavis.edu, mrdeng@ucdavis.edu, and zding@ucdavis.edu).}}

\markboth{Journal of \LaTeX\ Class Files,~Vol.~14, No.~8, August~2021}%
{Shell \MakeLowercase{\textit{et al.}}: A Sample Article Using IEEEtran.cls for IEEE Journals}


\maketitle

\begin{abstract}
Signal processing over single-layer graphs has become a mainstream 
tool owing to its power in revealing obscure underlying structures within data signals. 
However, many real-life datasets and systems, {including those in Internet of Things (IoT)}, are characterized by 
more complex interactions among distinct entities, which may represent multi-level interactions
that are harder to be captured with a single-layer graph, and can 
be better characterized by multilayers graph connections.
{Such 
multilayer or multi-level data structure can be 
more naturally modeled 
by high-dimensional multilayer graphs (MLG)}. 
To generalize traditional graph signal processing (GSP) over multilayer graphs for analyzing multi-level signal features 
and their interactions, 
this work proposes a tensor-based framework of 
{multilayer graph} signal processing (M-GSP).
Specifically,  
we introduce core concepts of M-GSP and 
study properties of MLG spectrum space, followed by fundamentals of {MLG-based} filter design.
To illustrate novel aspects of M-GSP,
we further explore its 
link with traditional
signal processing and GSP.
We provide example applications 
to demonstrate the efficacy
and benefits of applying {multilayer graphs} and M-GSP in
practical scenarios.
\end{abstract}


\section{Introduction}
\IEEEPARstart{G}{eometric} signal processing tools have found broad applications in data analysis to uncover obscure or hidden structures from complex datasets \cite{c1}. 
Various data sources, {such} as social networks, {Internet of Things (IoT) intelligence}, traffic flows and biological images,
often feature complex structures that pose challenges 
to traditional signal processing tools. Recently,
graph signal processing (GSP) emerges as an effective tool over the graph signal representation
\cite{c3}. For a signal with $N$ samples, a graph of $N$ nodes can be formed 
to model their underlying interactions \cite{c2}. In GSP, a graph Fourier space is also 
defined from the spectrum space of the representing matrix (adjacency/Laplacian) 
for signal processing tasks \cite{c4}, such as denoising \cite{c5}, resampling \cite{c6}, and classification \cite{c7}. Generalization of the more traditional GSP
includes signal processing over hypergraphs \cite{c8} and simplicial complexes \cite{c9},
{which are} suitable to model high-degree multi-lateral node relationships.

\begin{figure*}[htbp]
	\centering
	\subfigure[]{
		\label{ex1}
		\includegraphics[height=2.8cm]{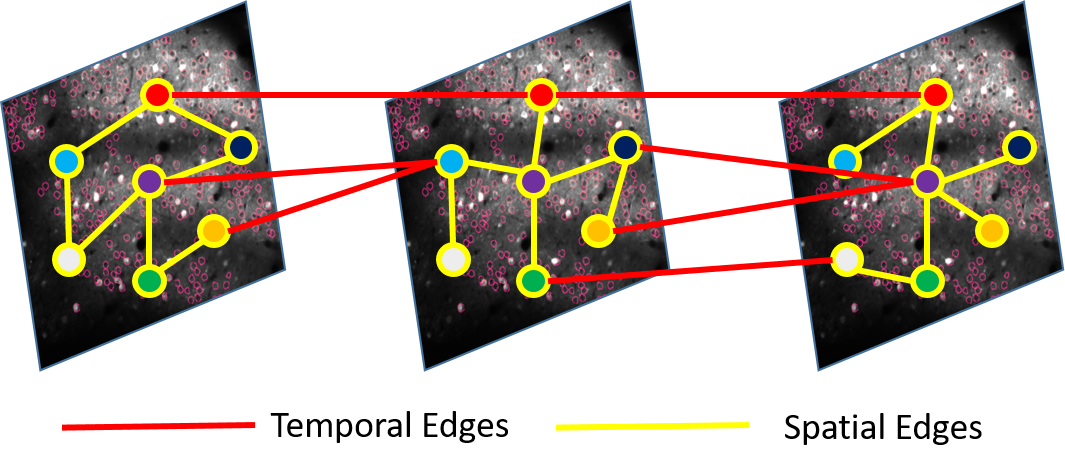}}
	\hspace{2.5cm}
	\subfigure[]{
		\label{ex2}
		\includegraphics[height=2.8cm]{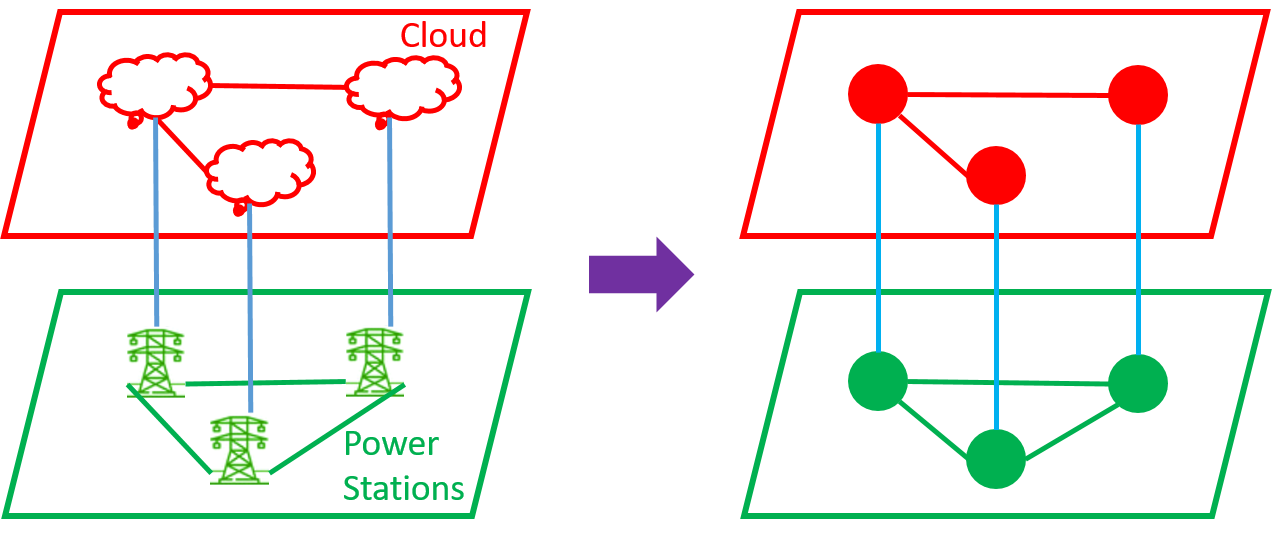}}
	\caption{Multilayer Graphs and Applications: (a) Video: each layer represents one frame of the video and the edges capture the spatial-temporal relationships; (b) Cyber-Physical System (CPS): each layer represents one component in CPS and the edges capture the physical connections.}
	\label{ex_mln}
\end{figure*}

Traditional graph signal processing tools generally describe
signals as graph nodes connected by one type of
edges. However, real-life systems and datasets may feature 
multi-facet interactions \cite{c11}. For example, in a video 
dataset modeled by spatial-temporal graph shown in Fig. \ref{ex1}, 
the nodes may exhibit different types of spatial connections 
at different temporal steps. It is harder
for single-layer graphs to model such multi-facet connections.
To model multiple layers of signal connectivity, we explore
a high-dimensional graph representation known as {multilayer graphs (MLG)} \cite{c10}.

{Multilayer graph, also named as multilayer network,} is a geometric model containing correlated layers with 
different structures and physical meanings, unlike
traditional single-layer graphs \cite{c11}. 
A typical example is smart grid consisting of two layers 
shown as Fig. \ref{ex2}: the power grid and the computation network. These two layers have different physical connectivity and rules \cite{c12}.
Still, signal interactions across the multiple layers in {MLG}
can be strongly correlated. Thus, separate representations by multiple single layer graphs
may fail to capture such characteristics. 
Consider a network consisting of a physical power layer and a cyber layer, 
the failure of one layer could trigger the failure of the other \cite{c13}. 
One example was the power line damage caused by a storm on September 28th of 2003.
Not only did it lead to the failure of several power stations, but also
disrupted communications as a result of power station breakdowns
that eventually affected 56 million people in Europe \cite{c14}. 

The complexity and multi-level interactions of MLG make the data reside on the irregular and high-dimensional structures, which do not directly lend themselves to standard GSP tools. For example, even though one can represent MLG by a supra-graph unfolding all the layers \cite{d1}, traditional GSP would treat interlayer and intralayer interactions equivalently in one spectrum space without differentiating the spectra of interlayer and intralayer signal correlations. 
Recently, there has been
growing interest in developing advanced GSP tools to process such multi-level structures. 
{In \cite{c19}, a joint time-vertex Fourier transform (JFT) is defined to process spatial-temporal graphs by applying graph Fourier transform (GFT) and discrete Fourier transform (DFT). Although JFT can process time-varying datasets, it does not provide more general temporal (interlayer) connectivity in a generic multilayer graph.
For flexible interlayer structure, a two-way GFT proposed in \cite{c16}, defines different graph Fourier spaces for interlayer and intralayer connections, respectively. However, its spatial interactions all lie within a single graph structure, thereby limiting the generalization of MLG. Expanding the works of \cite{c16}, the tensor-based multi-way graph signal processing framework (MWGSP) of \cite{c18}
relies on product graph. In MWGSP, separate factor graphs are constructed for each mode of a tensor-represented signal, and a joint spectrum combines factor graph spectra. However,
both JFT and MWGSP use the same intralayer connections for all layers. They do not admit different layers with heterogeneous graph structures, thereby limiting the ability to represent a general MLG.
}

Another challenge in MLG signal processing lies in 
the need for a suitable
mathematical representation. Traditional methods start with connectivity matrices. 
For example,  in \cite{d2}, a supra-adjacency matrix is defined to represent all layers equivalently while ignoring the natures of different layers.
One can also represent
each layer with an individual adjacency matrix \cite{c20}. 
However, such matrix-based representations mainly focus on the intralayer connections and lack representation for interlayer interactions. 
A more natural and general way may start with tensor representation 
\cite{c10}, which is particularly attractive in handling complex MLG graph analysis.

Our goal is to generalize graph signal processing for multilayer graphs
to model, analyze, and process signals
based on the {\em intralayer} and {\em interlayer} signal interactions.
To address the aforementioned challenges and to advance MLG processing, 
we present a novel tensor framework for {multilayer graph signal processing (M-GSP)}. We summarize the main contributions of this work as follows:
\begin{itemize}
	\item Leveraging tensor representation of MLG, {we introduce M-GSP from a spatial perspective, in which MLG signals and shifting of signals are defined};
	\item {Taking a spectrum perspective, we  introduce new concepts of spectrum space and spectrum transform for MLG.} For interpretability of spectrum space, we analyze the resulting MLG spectral properties and their distinction from existing GSP tools.
	\item We also present fundamentals of filter design in M-GSP, and suggest several practical applications based on the proposed framework, {including those in IoT systems}.
\end{itemize}

{We organize the technical presentation as 
follows. 
Section \ref{prelim} first summarizes 
preliminaries of traditional GSP and tensor analysis, before 
presenting representations of multilayer graphs within M-GSP in Section. \ref{funda}.
We then introduce the 
fundamentals 
of M-GSP spatial and spectrum analysis in Section \ref{spa_sec} and Section \ref{mln_spec}, respectively. We next 
discuss MLG filter design in Section \ref{fter}. We develop the physical insights and
spectrum interpretation of M-GSP concepts in Section \ref{discus}. With the newly proposed M-GSP framework, we provide several example applications to demonstrate its potential in Section \ref{app}, before concluding this paper in Section \ref{con}.}

\section{Preliminaries} \label{prelim}
\subsection{Overview of Graph Signal Processing}
Signal processing on graphs \cite{c1,c2,c3} studies signals that are discrete in some
dimensions by representing the 
irregular signal structure using a graph $\mathcal{G}=\{\mathcal{V},\mathbf{F}\}$, where $\mathcal{V}=\{v_1,v_2,\cdots, v_N\}$ is a set of $N$ nodes, and $\mathbf{F}\in\mathbb{R}^{N\times N}$ is the representing matrix (e.g., adjacency/Laplacian) describing the geometric structure of the graph $\mathcal{G}$. Graph signals are the attributes of nodes that {underlie} the graph structure. A graph signal can be written as vector
$\mathbf{s}=[s_1, s_2,\cdots, s_N]^\mathrm{T}\in\mathbb{R}^{N}$ where the superscript $(\cdot)^\mathrm{T}$ denotes matrix/vector transpose. 

With a graph representation $\mathbf{F}$ and a signal vector $\mathbf{s}$, 
the basic graph filtering (shifting) is defined via
\begin{equation}
	\mathbf{s}'=\mathbf{Fs}.
\end{equation}

The graph spectrum space, also known as the graph Fourier space is defined based on the eigenspace of the representing matrix. Let the eigen-decomposition of $\mathbf{F}$ 
be given by
$\mathbf{F}=\mathbf{V}\mathbf{\Lambda}\mathbf{V}^{-1}$,
where $\mathbf{V}$ is the matrix with eigenvectors of $\mathbf{F}$ as columns, and
diagonal matrix $\mathbf{\Lambda}$ consists of the corresponding eigenvalues. 
The graph Fourier transform (GFT) is defined as
\begin{equation}\label{GFT}
\hat{\mathbf{s}}=\mathbf{V}^{-1}\mathbf{s},
\end{equation}
whereas the inverse GFT is given by
$\mathbf{s}=\mathbf{V}\hat{\mathbf{s}}$.

From definitions of GFT, other concepts, such as
sampling theory \cite{c23}, filter design \cite{c24}, and frequency analysis \cite{c4} 
can be developed for signal processing and data analysis tasks.

\subsection{Introduction of Tensor Basics}
Before introducing the fundamentals of M-GSP, we first 
review some basics on tensors that are
useful for multilayer graph analysis.
Tensors can be viewed as multi-dimensional
arrays. The order of a tensor is the number of indices needed
to label a component of that array \cite{c25}.
For example, a third-order tensor has three indices. More specially, a scalar is a zeroth-order tensor; a vector is a first-order tensor; a matrix is a
second-order tensor; and an $M$-dimensional array is an $M$th-order tensor. 
For convenience, we use bold letters to represent the tensors excluding scalars, i.e.,  $\mathbf{A}\in\mathbb{R}^{I_1\times I_2\cdots\times I_N}$ represents an $N$th-order tensor with $I_k$ being the dimension of the $k$th order, and use $A_{{i_1}\cdots i_N}$ to represent the entry of $\mathbf{A}$ at position $(i_1,i_2,\cdots,i_N)$ 
with $1\leq i_k\leq I_k$ in this work. If $\mathbf{A}_f$ has a subscript $f$, we use $[A_f]_{{i_1}\cdots i_N}$ to denote its entries.

We now start with some useful definitions and tensor operations  
for the M-GSP framework \cite{c25}.
\subsubsection{Super-diagonal Tensor}
An $N$th-order tensor $\mathbf{A}\in\mathbb{R}^{I_1\times I_2\cdots\times I_N}$ is \textit{super-diagonal} if its entries $A_{i_1i_2\cdots i_N}\neq 0$ only for $i_1=i_2=\cdots=i_N$. 

\subsubsection{Symmetric Tensor}
A tensor is \textit{super-symmetric} if its elements remain constant under 
index permutation.
For example, a third-order $\mathbf{A}\in \mathbb{R}^{I\times I\times I}$ is \textit{super-symmetric} if
$A_{ijk}=A_{jik}=A_{kij}=A_{kji}=A_{jik}=A_{jki}$.
In addition, tensors can be \textit{partially symmetric} in two or more modes as well. For example, a third-order tensor $\mathbf{A}\in\mathbb{R}^{I\times I\times J}$ is \textit{symmetric} in the order one and two if
$A_{ijk}=A_{jik}$,
for $1\leq i,j\leq I$ and $1\leq k\leq J$.

\subsubsection{Tensor Outer Product}
The \textit{tensor outer product} between a $P$th-order tensor $\mathbf{U}\in \mathbb{R}^{I_1\times \cdots\times I_P}$ with entries $U_{i_1 ... i_P}$ and a $Q$th-order tensor $\mathbf{V} \in\mathbb{R}^{J_1\times \cdots\times J_Q}$ with entries $V_{j_1 ... j_Q}$ is denoted by 
\begin{equation}
\mathbf{W}=\mathbf{U} \circ \mathbf{V}. 
\end{equation}
The result $\mathbf{W}\in \mathbb{R}^{I_1\times \cdots\times I_P\times J_1\times \cdots\times J_Q}
$ is a $(P+Q)$th-order tensor, whose entries are calculated by
\begin{equation}
W_{i_1 ... i_P j_1 ... j_Q}= U_{i_1 ... i_P} \cdot V_{j_1 ... j_Q}.
\end{equation}
The tensor outer product is useful to construct a higher order tensor from several lower order tensors.

\subsubsection{n-mode Product}
The \textit{n-mode product} between a tensor $\mathbf{U}\in \mathbb{R}^{I_1\times \cdots \times I_P}$ and a matrix $\mathbf{V}\in \mathbb{R}^{J\times I_n}$ is denoted by 
\begin{equation}
\mathbf{W}=\mathbf{U}\times_n \mathbf{V}\in \mathbb{R}^{I_1\times \cdots\times
	I_{n-1}\times J \times I_{n+1}\times \cdots \times I_P}.
\end{equation}
Each element in $\mathbf{W}$ is given by
\begin{equation}
	W_{i_1 i_2 \cdots i_{n-1} j i_{n+1} \cdots i_P}=\sum_{i_n=1}^{I_n}U_{i_1\cdots i_P}V_{j i_n}.
\end{equation}
Note that the $n$-mode product is a different operation from matrix product.

\begin{figure*}[t]
	\centering
	\subfigure[CP Decomposition.]{
		\label{cp}
		\includegraphics[height=2cm]{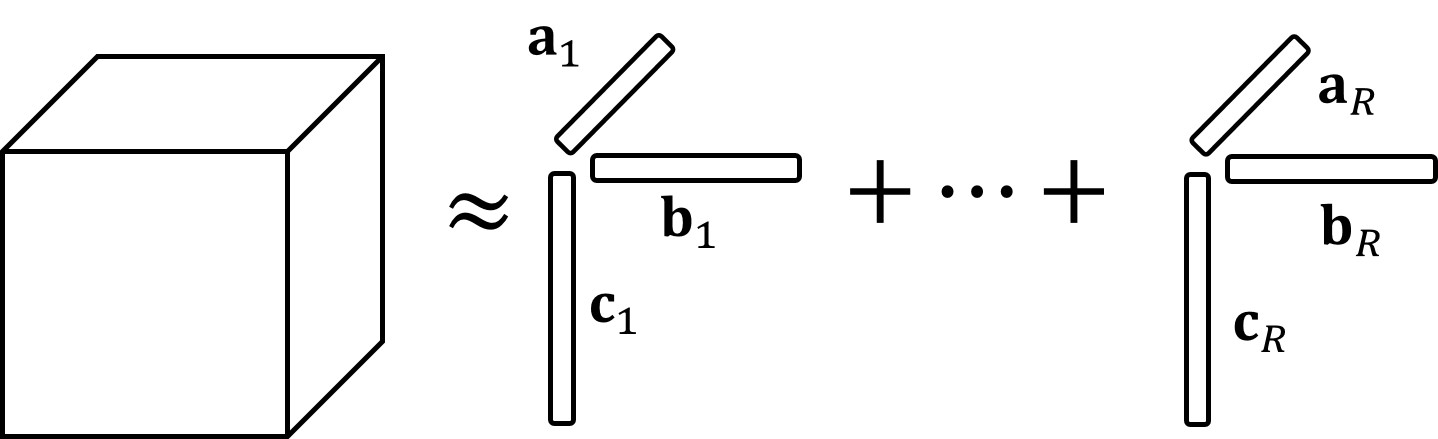}}
	\hspace{2cm}
	\subfigure[Tucker Decomposition.]{
		\label{tk}
		\includegraphics[height=2.5cm]{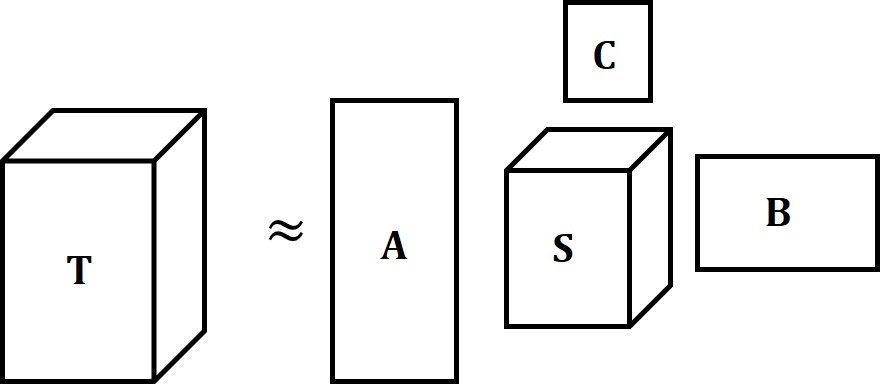}}
	\caption{Diagram of Tensor Decompositions.}
	\label{td}
\end{figure*}

\subsubsection{Tensor Contraction}
In M-GSP, the contraction (inner product) between a forth order tensor $\mathbf{A}\in\mathbb{R}^{M\times 
	N\times M \times N}$ and a matrix $\mathbf{x} \in \mathbb{R}^{M\times N}$ in the third and forth order is defined as
\begin{equation}\label{contract}
\mathbf{y}=\mathbf{A}\diamond \mathbf{x}\in \mathbb{R}^{M\times N},
\end{equation}
where $y_{\alpha i}=\sum_{\beta=1}^M\sum_{j=1}^N A_{\alpha i \beta j}x_{\beta j}$.

In addition, the contraction between two fourth-order 
tensor $\mathbf{U},\mathbf{V}\in\mathbb{R}^{M\times 
	N\times M \times N}$
is defined as 
\begin{equation}\label{oodot}
\mathbf{W}=\mathbf{U}\odot \mathbf{V}\in\mathbb{R}^{M\times 
	N\times M \times N},
\end{equation}
whose entries are $W_{\alpha i \epsilon p}=\sum_{\beta j} U_{\alpha i \beta j} V_{\beta j \epsilon p}$.

\subsubsection{Tensor Decomposition}
Tensor decompositions are useful tools to extract the underlying information of tensors. Particularly,
CANDECOMP/PARAFAC (CP) decomposition decomposes a tensor as a sum of {the tensor outer product of} rank-one tensors \cite{c25,c26}.
{For example, a third order tensor $\mathbf{T}\in\mathbb{R}^{I\times
	J\times K}$ 
is decomposed by CP decomposition into
\begin{equation}
\mathbf{T}\approx\sum_{r=1}^{R}\mathbf{a}_r\circ \mathbf{b}_r\circ \mathbf{c}_r,\quad
\mathbf{a}_r\in \mathbb{R}^I, \mathbf{b}_r\in \mathbb{R}^J, \mathbf{c}_r\in \mathbb{R}^K
\end{equation}
where integer $R$ 
denotes the
rank, i.e., the lowest number of rank-one tensors in the decomposition. We illustrate CP decomposition for a third-order tensor in Fig. \ref{cp}, 
which could be viewed as factorization of the tensor.}

Another important decomposition is the Tucker decomposition, which is in the form of higher-order PCA.
More specifically, Tucker decomposition decomposes a tensor
into a core tensor multiplied by a matrix along each mode \cite{c25}. 
{Defining a  core tensor $\mathbf{S}=[S_{pqr}]\in\mathbb{R}^{P
	\times Q \times R}$.
Defining
$\mathbf{A}\in\mathbb{R}^{I\times P}, \mathbf{B}\in\mathbb{R}^{J\times Q}, \mathbf{C}\in\mathbb{R}^{K\times R}$, the Tucker decomposition of a third-order tensor $\mathbf{T}\in\mathbb{R}^{I\times
	J\times K}$ is
\begin{align}
\mathbf{T}&\approx\mathbf{S}\times_1 \mathbf{A}\times_2 \mathbf{B}\times_3 \mathbf{C},
\quad
\nonumber\\
&=\sum_{p=1}^{P}\sum_{q=1}^{Q}\sum_{r=1}^{R} S_{pqr}\mathbf{a}_p\circ \mathbf{b}_q\circ \mathbf{c}_r,
\end{align}
where $\mathbf{A}=[\mathbf{a}_1\; \cdots\; \mathbf{a}_p]$, 
$\mathbf{B}=[\mathbf{b}_1\; \cdots\; \mathbf{b}_Q]$, $\mathbf{C}=[\mathbf{c}_1\; \cdots\; \mathbf{c}_R]$. The Tucker decomposition for a third-order tensor is illustrated in Fig. \ref{tk}.
Tucker decomposition reduces to CP decomposition if the core tensor is limited to be super-diagonal.}

Other typical decompositions include Higher-Order SVD (HOSVD) \cite{c27}, orthogonal CP-decomposition 
\cite{c28}, and Tensor-Train decomposition \cite{c29}.
Interested readers are referred to the tutorial \cite{c25} for more 
details.

\section{{Representations of MLG in M-GSP}} \label{funda}
{In this section, we introduce the fundamental representations of MLG within the M-GSP framework.}
\begin{figure}[t]
	\centering
	\subfigure[]{
		\label{mln1}
		\includegraphics[height=3.5cm]{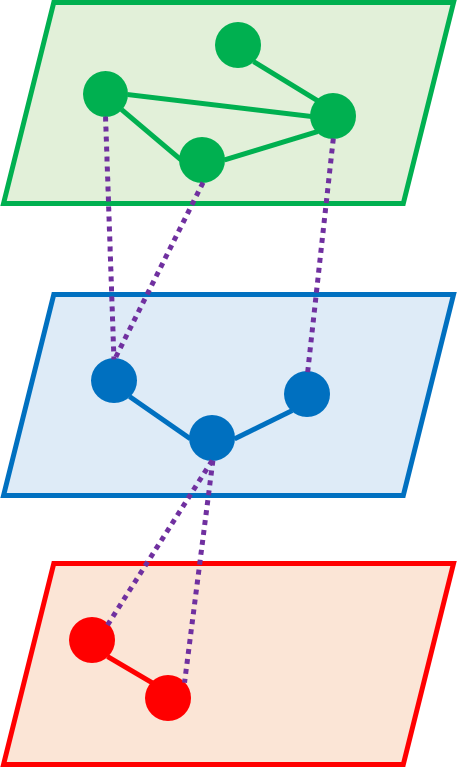}}
	\hspace{2cm}
	\subfigure[]{
		\label{mlp1}
		\includegraphics[height=3.5cm]{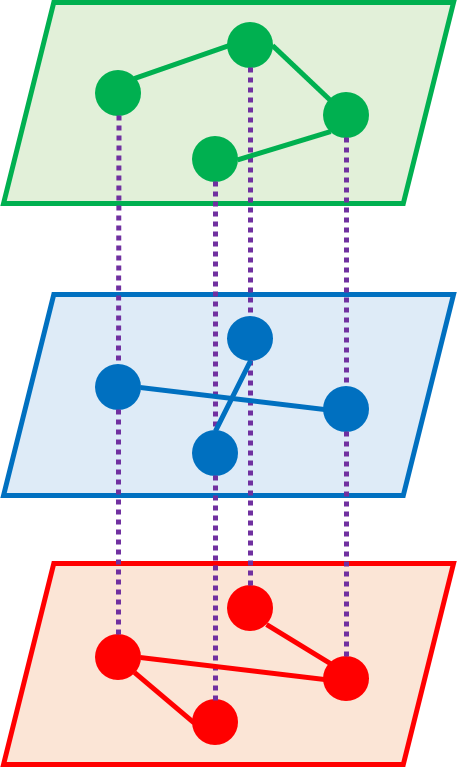}}
	\caption{Example of multilayer graphs: (a) A three-layer interconnected graph; (b) A three-layer multiplex graph.}
	\label{ex_mln1}
\end{figure}

\subsection{Multilayer Graphs}
{Multilayer graph, also referred to as multilayer network, is an important geometric model in complex systems \cite{c11}. In this work, we refrain from using the ``network" terminology because of its diverse meanings in various field
ranging from communication networking to deep learning.
From here onward, unless otherwise specified, we shall use the
less ambiguous term of multilayer graph.}

We first provide definitions of multilayer graphs (MLG).
\begin{definition}[Multilayer Graph]
	A multilayer graph with $K$ nodes and $M$ layers is defined as $\mathcal{M}=\{\mathcal{V},\mathcal{L},\mathbf{F}\}$, where $\mathcal{V}=\{v_1,v_2,\cdots,v_K\}$ is the set of nodes, $\mathcal{L}=\{l_1,l_2,\cdots,l_M\}$ denotes the set of layers with each layer $l_i=\{v_{i_1},\cdots,v_{i_n}\}$ being the subsets of $\mathcal{V}$, whereas $\mathbf{F}$ is the algebraic representation describing node interactions.
\end{definition}
Note that, we mainly focus on the layer-disjoint multilayer graph \cite{c11} where each node exists exactly in one layer, since layers denote different phenomena. For example, in a smart grid, a station with functions in both power grid and communication network, is usually modeled as two nodes in a two-layer graph for the network analysis \cite{d3}.

In multilayer graphs, edges connect nodes in the same layer (intralayer edges) or nodes of different layers (interlayer edges) \cite{c30}. There are two main types of multilayer graphs: \textit{multiplex graph} and \textit{interconnected graph} \cite{c31}. In a \textit{multiplex graph}, each layer has the same number of nodes, and each node only connects with
their 1-to-1
matching counterparts in other layers to form interlayer connections. Typically, multiplex graphs characterize different 
types of interactions among the same (or a similar) set of physical entities. For example, the spatial-temporal connections among a set of nodes can be intuitively modeled as a multiplex graph \cite{c31}. In the \textit{interconnected graphs}, each layer may have
different numbers of nodes without a 1-to-1 counterpart.
Their interlayer connections could be more flexible.
Examples of a three-layer multiplex graph and a three-layer interconnected graph are shown in Fig. \ref{ex_mln1}, where different colors represent different layers, solid lines represent intralayer connections, and dash lines indicate interlayer connections.

\subsection{Algebraic Representation}
To capture the high-dimensional `multilayer' interactions between different nodes, we use tensor as algebraic representation of MLG for the proposed M-GSP framework \cite{c10}.

\subsubsection{MLG with same number of nodes in each layer}
To better interpret the tensor representation of a multilayer graph, we start from a simpler type of MLG, 
in which each layer contains the same number of nodes.
For a multilayer graph $\mathcal{M}=\{\mathcal{V},\mathcal{L}\}$ with $|\mathcal{L}|=M$ layers and $N$ nodes in each layer, i.e., $|l_i|=N$ for $1\leq i \leq M$, it could be interpreted as embedding the interactions between a set of $N$ `entities' (not nodes) into a set of $M$ layers. The nodes in different layers can be viewed as the projections of the entities.
For example, the video datasets could be modeled by the spatial connections between objects (entities) into different temporal frames (layers).

Mathematically, the process of
embedding (projecting) entities can be viewed as a tensor product, and 
graph connections can be represented by tensors \cite{c10}.
For {convenience}, we use Greek letters $\alpha,\beta,\cdots$ to indicate each layer
and Latin letters $i,j,\cdots$ to indicate each interpretable `entity'
with corresponding node in each layer. 
Given a set of entities $\mathcal{X}=\{x_1,x_2,\cdots,x_N\}$, one can construct a vector $\mathbf{e}_i
=[0,\cdots,0,1,0,\cdots,0]^\mathrm{T}$ whose sole
nonzero element is 
its $i-$th element (equal to 1)
to characterize each entity $i$.
Thus, interactions of two entities can be represented by a second-order tensor $\mathbf{A}_X=\sum_{i,j=1}^N a_{ij} \mathbf{e}_i \circ \mathbf{e}_j\in\mathbb{R}^{N\times N}$, where $a_{ij}$ is the intensity of the relationship between entity $i$ and $j$.
Similarly, given a set of layers $\mathcal{L}=\{l_1,l_2,\cdots,l_M\}$, a vector $\mathbf{e}_{\alpha}\in \mathbb{R}^{M}$ can 
capture the properties of the layer $\alpha$, and the connectivity between two layers could be represented by $\mathbf{A}_L=\sum_{\alpha,\beta=1}^M b_{\alpha\beta}\mathbf{e}_\alpha \circ \mathbf{e}_\beta \in \mathbb{R}^{M\times M}$. 
Following this approach,
connectivity between the projected nodes of the entities
in the layers can be represented by a {fourth-order} tensor
\begin{equation}\label{tensor_construction}
\mathbf{A}=\sum_{\alpha,\beta=1}^{M}\sum_{i,j=1}^N w_{\alpha i \beta j}\mathbf{e}_\alpha \circ \mathbf{e}_i\circ \mathbf{e}_\beta\circ\mathbf{e}_j \in \mathbb{R}^{M\times N\times M \times N},
\end{equation}
where $w_{\alpha i \beta j}$ is the weight of connection between the entity $i$'s projected node on layer $\alpha$ and the entity $j$'s projected node on layer $\beta$. More specially, 
the {fourth-order} tensor becomes the adjacency tensor of the multilayer graph, where each entry $A_{\alpha i \beta j}=w_{\alpha i \beta j}$ characterizes the edge between the entity $i$'s projected node on layer $\alpha$ and the entity $j$'s projected node on layer $\beta$. 
Thus, similar to the adjacency matrix whose 2-D entries indicate
whether and how two nodes are pairwise connected by a
simple edge in the normal graphs, we adopt an adjacency tensor $\mathbf{A}\in\mathbb{R}^{M\times N\times M \times N}$ to represent the multilayer graph with the same number of nodes 
in each layer as follows.
\begin{definition}[Adjacency Tensor]
	A multilayer graph $\mathcal{M}$, with $|\mathcal{L}|=M$ layers and $|l_i|=N$ nodes in each layer $i$, can be represented by a {fourth-order} adjacency tensor $\mathbf{A}\in \mathbb{R}^{M\times 
		N\times M \times  N}$ defined as
	\begin{equation}
	\mathbf{A}=(A_{\alpha i \beta j}), \quad 1\leq \alpha,\beta\leq M, 1\leq i,j\leq N.
	\end{equation}
	Here, each entry $A_{\alpha i \beta j}$ of the adjacency tensor $\mathbf{A}$ indicates the intensity of the edge between the entity $j$'s projected node on layer $\beta$ and entity $i$'s projected node on layer $\alpha$.
\end{definition}

Clearly, for a single layer graph, $\mathbf{e}_\alpha$ is a 
scalar $1$ and the {fourth-order} tensor degenerates to
the adjacency matrix of a normal graph. 
Similar to $A_{ij}$ in an adjacency matrix which indicates the direction from the node $v_j$ to $v_i$,  $A_{\alpha i \beta j}$ also indicates the direction from the node $v_{\beta j}$ to the node $v_{\alpha i}$ in a MLG.
Note that, vectors $\mathbf{e}_i$ and $\mathbf{e_\alpha}$ are not eigenvectors of the adjacency tensor. They are merely the vectors characterizing
features of the entities and layers, respectively. 
We 
shall discuss the MLG-based spectrum space in Section \ref{mln_spec}. 

Given an adjacency tensor, we can
define the Laplacian tensor of the multilayer graph similar to that in a single-layer graph. Denoting the degree (or multi-strength) of the entity's $i$'s projected node $v_{\alpha i}$ on layer $\alpha$ as $d(v_{\alpha i})$ which is a summation over weights of different
natures (inter- and intra- layer edges), the degree tensor $\mathbf{D}\in \mathbb{R}^{M\times N\times 
	M \times N}$ is defined as a diagonal tensor with entries $D_{\alpha i \alpha i}=d(v_{\alpha i})$ for $1\leq i\leq N, 1\leq \alpha \leq M$, whereas its other entries are zero. The Laplacian tensor can be defined as follows.
\begin{definition}[Laplacian Tensor]
	A multilayer graph $\mathcal{M}$, with $|\mathcal{L}|=M$ layers and $|l_i|=N$ nodes in each layer $i$, can be represented by a {fourth-order} Laplacian tensor $\mathbf{L}\in \mathbb{R}^{M\times N\times 
		M \times N} $ defined as
	$\mathbf{L=D-A}$,
	where $\mathbf{A}$ is the adjacency tensor and $\mathbf{D}$ is the degree tensor.
\end{definition}
The Laplacian tensor can be useful to analyze propagation processes 
such as
diffusion or random walk \cite{c10}. Both adjacency and Laplacian 
tensors are important algebraic representations of the MLG depending on datasets and user objectives. {For convenience, we use a tensor $\mathbf{F}\in \mathbb{R}^{M\times N\times 
		M \times N}$ as a general representation of a given MLG $\mathcal{M}$. As the adjacency tensor is more general, the
	representing tensor \textbf{F} refers to the adjacency tensor in this paper unless specified otherwise.}

\subsubsection{Representation of General MLG}
Representing a general multilayer graph with different number of nodes in each layer always remains a challenge if one aims to distinguish the interlayer and intralayer connection features. In JFT \cite{c19} and MWGSP \cite{c18}, all layers must reside on the same underlying graph structure which restrict the number of nodes to be the same in each layer. Similarly, a reconstruction is also needed to represent a general MLG by the forth-order tensor in M-GSP.
Note that, although M-GSP also needs a reconstruction to represent a general MLG, we allow different layers with heterogeneous graph structures, which provides additional flexibility than JFT and MWGSP.

There are mainly two ways to reconstruct: 1) Add isolated nodes to 
layers with fewer nodes to reach $N$ nodes \cite{c12} and set the augmented signals as zero; and 2) Aggregate several nodes into super-nodes for 
layers with $|l_i|>N$ \cite{c33} and merge the corresponding signals.
Since 
isolated nodes do not interact with any other nodes, it does not change the topological structure of the original multilayer architecture in the sense of signal shifting while the corresponding spectrum space could still be changed.
The aggregation method depends on how efficiently we can aggregate 
redundant or similar nodes. Different methods can be applied depending on specific tasks. For example, if one wants to explore the cascading failure in a physical system, the method based on isolated nodes is more suitable. For the applications, such as video analysis where pixels can be intuitively merged as superpixels, the aggregation method can be also practical.

In addition, although the {fourth-order} representing tensor can be viewed as the projection of several entities into different layers in Eq. (\ref{tensor_construction}), the entities and layers can be
virtual and not necessarily
physical to capture the underlying structures of the datasets.
The information within the multilayer graphs, together with definitions of the underlying virtual entities and layers, should only depend on the structure of the multilayer graphs. We will illustrate this further in Section \ref{exn_la}.

\subsection{Flattening and Analysis} \label{mln_flat}
In this part, we introduce the flattening of the multilayer graph, which could simplify some operations in the tensor-based M-GSP.
For a multilayer graph $\mathcal{M}=\{\mathcal{V},\mathcal{L},\mathbf{F}\}$ with $M$ layers and $N$ nodes in each layer, its {fourth-order} representing tensor $\mathbf{F}\in \mathbb{R}^{M\times N\times 
	M \times N}$ can be flattened into a second-order matrix to capture the overall edge weights. There are two main flattening schemes in the sense of entities and layers, respectively:
\begin{itemize}
	\item Layer-wise Flattening: The representing tensor $\mathbf{F}$ can be flattened into $\mathbf{F}_{FL} \in \mathbb{R}^{MN\times MN}$ with 
	each element 
	\begin{equation}\label{lwise}
	{[F_{FL}]}_{{N(\alpha-1)+i, N(\beta-1)+j}}=F_{\alpha i \beta j}.
	\end{equation}
	\item Entity-wise Flattening:  The representing tensor $\mathbf{F}$ can be flattened into $\mathbf{F}_{FN} \in \mathbb{R}^{NM\times NM}$ with each element 
	\begin{equation}\label{ewise}
	{[F_{FN}]}_{{M(i-1)+\alpha, M(j-1)+\beta}}=F_{\alpha i \beta j}.
	\end{equation}
\end{itemize}

\begin{figure}[t]
	\centering
	\includegraphics[width=3.5in]{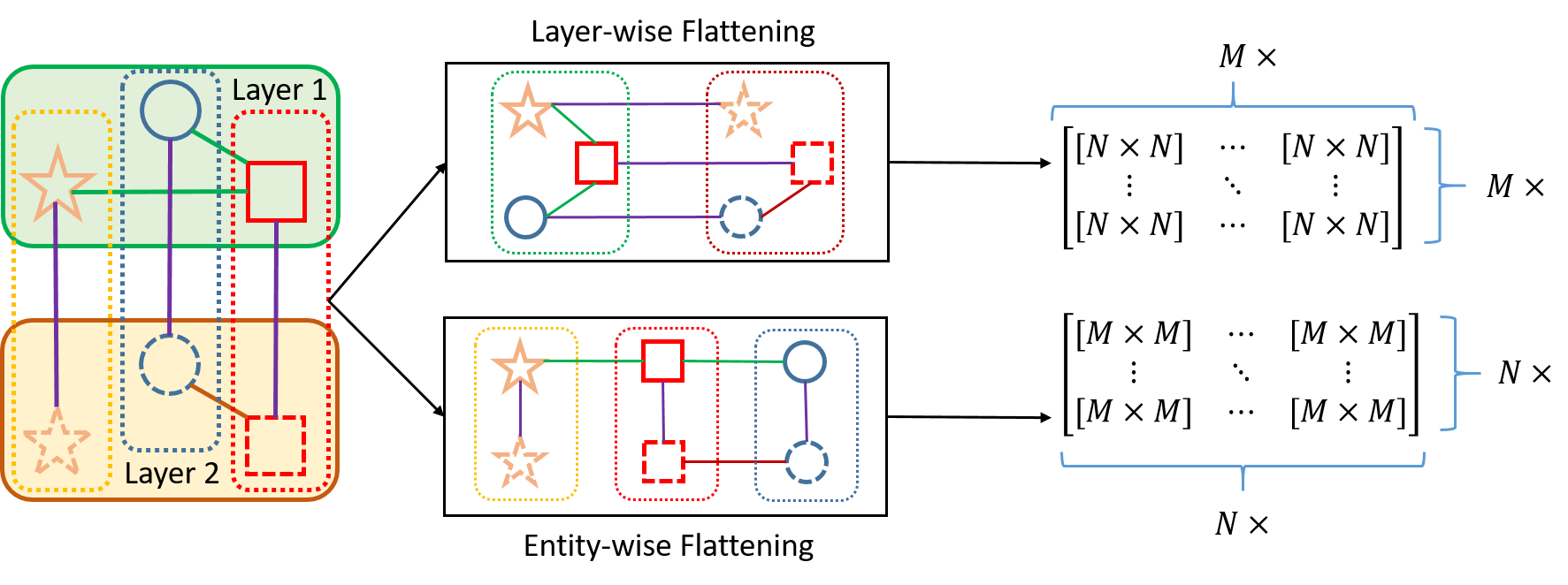}
	\caption{Example of Multilayer Graph Flattening.}
	\label{flat}
\end{figure}

These two flattening methods provide two ways to interpret the graph structure. In the first method, the flattened multilayer graph has $M$ clusters with $N$ nodes in each cluster. The nodes in the same cluster have the same function (belong to the same layer). In the second method, the flattened graph has $N$ clusters with $M$ nodes in each cluster. 
Here, the nodes in the same cluster are from the same entity. 
Examples of the tensor flattening of a two-layer graph with $3$ nodes in each layer are shown in Fig. \ref{flat}. From the examples, we can see that the diagonal blocks in $\mathbb{R}^{N\times N}$ are the intralayer connections for each layer and other blocks describe the interlayer connections through
{\em layer-wise flattening}; and the diagonal block in $\mathbb{R}^{M\times M}$ describe the `intra-entity' connections and other elements represent the `inter-entity' connections in  
{\em entity-wise flattening}. Although these two flattening schemes define the same MLG with a different
indexing of vertices, they are still helpful to analyze the MLG spectrum space.

\section{Spatial Definitions in M-GSP} \label{spa_sec}
Based on the tensor representation, we now define
signals and signal shifting over the multilayer graphs. In GSP, each signal sample is the attribute of one node. Typically, a graph signal can be represented by an $N$-length vector for a graph with $N$ nodes. Recall that in traditional GSP \cite{c2},  basic signal shifting is defined with the representing matrix as the shifting filter. Thus, in M-GSP, we can also define the signals and signal shifting based on the filter implementation.

In M-GSP, each signal sample is also related to one node in the multilayer graph. Intuitively, if there are $K=MN$ nodes, there are $MN$ signal samples in total. Similar to GSP, we use the representing (adjacency/Laplacian) tensor $\mathbf{F}\in \mathbb{R}^{M
	\times N\times 
	M \times N}$ as the basic MLG-filter. Since the input signal and the output signal of the MLG-filter should be consistent in the tensor size, we define a special form of M-GSP signals to work with the representing tensor as follows.

\begin{definition} [Signals over Multilayer Graphs]
	For a multilayer graph $\mathcal{M}=\{\mathcal{V},\mathcal{L},\mathbf{F}\}$, with $|\mathcal{L}|=M$ layers and $|l_i|=N$ nodes in each layer $i$, the definition of multilayer graph signals is a second-order tensor
	\begin{equation}
	\mathbf{s}=(s_{\alpha i})\in\mathbb{R}^{M\times
		N}, \quad 1\leq \alpha \leq M, 1\leq i\leq N,
	\end{equation}
	where the entry $s_{\alpha i}$ is the signal sample in the projected node of entity $i$ on layer $\alpha$. 
\end{definition}

Note that, if the multilayer graph degenerates to a single-layer graph with $M=1$, the multilayer graph signal becomes an $N$-length vector, which is similar to that in the traditional GSP. Similar to the representing tensor, the tensor signal $\mathbf{s}\in\mathbb{R}^{M\times N}$ can
also be flattened as a vector in $\mathbb{R}^{MN}$:
\begin{itemize}
	\item Layer-wise flattening: $\mathbf{s}_L\in \mathbb{R}^{MN}$ whose entries are calculated as
	$[s_L]_{N(\alpha-1)+i}=s_{\alpha i}$.	
	\item Entity-wise flattening: $\mathbf{s}_N\in \mathbb{R}^{NM}$ whose entries are calculated as
	$[s_N]_{M(i-1)+\alpha}=s_{\alpha i}$.	
\end{itemize}

Given the definitions of multilayer graph signals and filters, we now introduce the definitions of signal shifting in M-GSP. In traditional GSP, the signal shifting is defined as product between signal vectors and representing matrix. Similarly, we define the shifting in the multilayer graph based on the contraction (inner product) between the representing tensor and tensor signals.

\begin{definition}[Signal Shifting over Multilayer Graphs]
	Given the representing tensor $\mathbf{F}\in\mathbb{R}^{M\times 
		N\times M \times N}$ and the tensor signal $\mathbf{s}\in\mathbb{R}^{M\times N}$ defined over a multilayer graph $\mathcal{M}$, the signal shifting is defined as the contraction (inner product) between $\mathbf{F}$ and $\mathbf{s}$ in one entity-related order and one layer-related order, i.e.,
	\begin{equation} \label{shift_mln}
	\mathbf{s}'=\mathbf{F}\diamond \mathbf{s}\in \mathbb{R}^{M\times 
		N},
	\end{equation}
	where $\diamond$ is the contraction between $\mathbf{F}$ and $\mathbf{s}$ defined in Eq. (\ref{contract}).
\end{definition}
The elements in the shifted signal $\mathbf{s}'$ are calculated as
\begin{equation} \label{diffuse}
s'_{\alpha i}=\sum_{\beta=1}^M\sum_{j=1}^N F_{\alpha i \beta j}s_{\beta j}.
\end{equation}

\begin{figure}[t]
	\centering
	\includegraphics[width=1.4in]{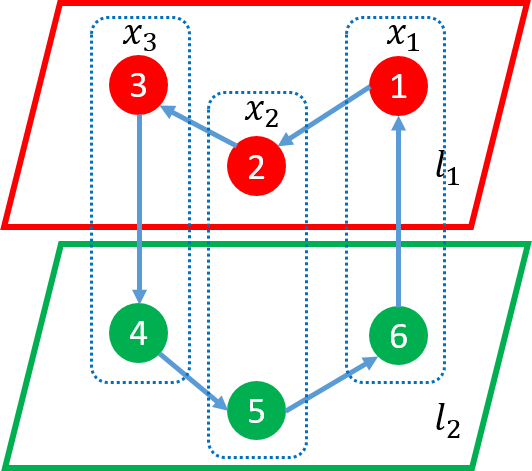}
	\caption{Example of Multilayer Graph Shifting.}
	\label{mln_shi}
\end{figure}

From Eq. (\ref{diffuse}), there are two important factors to construct the shifted signal: 1) The signal in the neighbors {(both intra- and inter- layer interactions)} of the node $v_{\alpha i}$; and 2) The intensity of interactions between the node $v_{\alpha i}$ and its neighbors. Then, the signal shifting is related to the diffusion process over the multilayer graphs. More specifically, if $\mathbf{F}$ is the adjacency tensor,
signals shift in directions of edges. To better illustrate the signal shifting based on the adjacency tensor, we use a two-layer directed network shown in Fig. \ref{mln_shi} as an example. In this multilayer graph, the original signal $\mathbf{s}$ is defined as

\begin{equation}
\mathbf{s}=\begin{bmatrix}
1 & 2 & 3\\
6 & 5 & 4
\end{bmatrix},
\end{equation}
and the adjacency tensor $\mathbf{A}$ is defined as
\begin{equation}\nonumber
\mathbf{A}_{(1,:,1,:)}=
\begin{bmatrix}
0 & 0 & 0\\
1 & 0 & 0\\
0 & 1 & 0
\end{bmatrix},
\mathbf{A}_{(2,:,2,:)}=
\begin{bmatrix}
0 & 1 & 0\\
0 & 0 & 1\\
0 & 0 & 0
\end{bmatrix},
\end{equation}
\begin{equation}\nonumber
\mathbf{A}_{(2,:,1,:)}=
\begin{bmatrix}
0 & 0 & 0\\
0 & 0 & 0\\
0 & 0 & 1
\end{bmatrix},
\mathbf{A}_{(1,:,2,:)}=
\begin{bmatrix}
1 & 0 & 0\\
0 & 0 & 0\\
0 & 0 & 0
\end{bmatrix},
\end{equation}
for each fiber. Then, the shifted signal is calculated as
\begin{equation}\label{res}
\mathbf{s}'=\begin{bmatrix}
6 & 1 & 2\\
5 & 4 & 3
\end{bmatrix}.
\end{equation}
From Eq. (\ref{res}), we can see that the signal shift one step following the direction of the links.

Meanwhile, if $\mathbf{F}$ is the Laplacian tensor, Eq. (\ref{diffuse}) can be written as 
\begin{equation}
s'_{\alpha i}=\sum_{\beta=1}^M\sum_{j=1}^N A_{\alpha i \beta j} (s_{\alpha i}-s_{\beta j}),
\end{equation}
which is the weighted average of difference with neighbors.

\section{Multilayer Graph spectrum space} \label{mln_spec}
In traditional GSP,  graph spectrum space is defined according to the eigenspace of the representing matrix \cite{c2}. Similarly, we define the MLG spectrum space based on the decomposition of the representing tensor.
Since tensor decomposition is less stable when exploring the factorization of 
a specific order or when
extracting the separate features in the asymmetric tensors, we will mainly focus on spectral properties of undirected multilayer graphs in 
this section for simplicity and
clarity of presentation. {For directed MLG, we provide alternative spectrum definitions in the Appendix and will explore detailed analysis 
in future works.}

\subsection{Joint Spectral Analysis in M-GSP} \label{JSA}
For a multilayer graph $\mathcal{M}=\{\mathcal{V},\mathcal{L},\mathbf{F}\}$ with $M$ layers and $N$ nodes,
the eigen-tensor $\mathbf{V}\in\mathbb{R}^{M\times N}$ of the representing tensor $\mathbf{F}$ is defined in the tensor-based multilayer graph theory \cite{c10} as 
\begin{equation}
   \mathbf{F}\diamond \mathbf{V}=\lambda \mathbf{V}. 
\end{equation}
More specifically, $\mathbf{F}\in\mathbb{R}^{M\times N\times M\times N}$ can be decomposed as 
\begin{align}\label{ted}
\mathbf{F}
&=\sum_{k=1}^{MN}\lambda_k \mathbf{V}_k \circ\mathbf{V}_k\\
&=\sum_{\alpha=1}^M\sum_{i=1}^N \lambda_{\alpha i} \mathbf{V}_{\alpha i} \circ\mathbf{V}_{\alpha i},
\end{align}
where $\lambda_k$ is the eigenvalues and $\mathbf{V}_k\in\mathbb{R}^{M\times N}$ is the corresponding eigen-tensor. Note that $\mathbf{V}_{\alpha i}$ just relabels the index of $\mathbf{V}_k$, and there is no specific order for $\mathbf{V}_{\alpha i}$ here.

Similar to the traditional GSP where the graph Fourier space is defined by the eigenvectors of the representing matrix, we define the joint MLG Fourier space as follows.
\begin{definition}[Joint Multilayer Graph Fourier Space]
	For a multilayer graph $\mathcal{M}=\{\mathcal{V},\mathcal{L},\mathbf{F}\}$ with $M$ layers and $N$ nodes, the MLG Fourier space is defined as the space consisting of all spectral tensors $\{\mathbf{V}_1,\cdots,\mathbf{V}_{MN}\}$, which characterizes the joint features of entities and layers.
\end{definition}

Recall that in GSP, the GFT is defined based on the inner product of $\mathbf{V}^{-1}$ and the signals $\mathbf{s}$ defined in Eq. (\ref{GFT}). Similarly, we can define the M-GFT based on the spectral tensors of the representing tensor $\mathbf{F}$ to capture joint features of inter- and intra- layer interactions as follows.
\begin{definition}[Joint M-GFT]
	Let $\mathbf{U}_\mathcal{F}=(\mathbf{V}_{\alpha i})\in\mathbb{R}^{M\times N \times M \times N}$ consist of spectral tensors of the representing tensor $\mathbf{F}$, where
	$[U_\mathcal{F}]_{\alpha i \beta j}=[V_{\alpha i}]_{\beta j}$.
	
	The {joint M-GFT (M-JGFT)} can be defined as the contraction between $\mathbf{U}_\mathcal{F}$ and the tensor signal $\mathbf{s}\in\mathbb{R}^{M\times N}$, i.e.,
	\begin{equation}\label{M-GFT}
	\hat{\mathbf{s}}=\mathbf{U}_\mathcal{F} \diamond \mathbf{s}.
	\end{equation}
\end{definition}

Now, we show how to obtain the eigen-tensors. Implementing the flattening
analysis, we have the following properties.

\begin{property} 
	The two types of flattened tensor in Eq. (\ref{lwise}) and Eq. (\ref{ewise}) lead to the same eigenvalues.
\end{property}
\begin{proof}
	Suppose $(\lambda,\mathbf{x})$ is an eigenpair of $\mathbf{A}_{FL}$, i.e., 
	\begin{equation}
	\mathbf{A}_{FL}\cdot\mathbf{x}=\lambda\mathbf{x}.
	\end{equation}
	Let $x_{N(\alpha-1)+i}=y_{M(i-1)+\alpha}$. Since
	\begin{align}
	A_{\alpha i \beta j}&=
	[A_{FL}]_{N(\alpha-1)+i,N(\beta-1)+j}\nonumber\\
	&=[A_{FN}]_{M(i-1)+\alpha,M(j-1)+\beta},
	\end{align}
	we have 
	\begin{equation}
	\mathbf{A}_{FN}\cdot\mathbf{y}=\lambda\mathbf{y}.
	\end{equation}
	Thus, $\lambda$ is also an eigenvalue of $\mathbf{A}_{FN}$.
\end{proof}
This property shows that the flattened tensors are the {reshaped original representing tensor}, and could capture some of the spectral properties as follows.

\begin{property}\label{eig}
	Given the eigenpair $(\lambda_{FL}, \mathbf{x})$ of the layer-wise flattened tensors, the eigenpair $(\lambda,\mathbf{V})$ of the original representing tensor can be calculated as
	$\lambda=\lambda_{FL}$, and 
	$V_{\alpha i}=x_{N(\alpha-1)+i}$.
	Similarly, given the eigenpair $(\lambda_{FN}, \mathbf{y})$ of the entity-wise flattened tensor,
	the eigenpair $(\lambda,\mathbf{V})$ of the original representing tensor can be calculated as
	$\lambda=\lambda_{FN}$, and 
	$V_{\alpha i}=y_{M(i-1)+\alpha}$.
\end{property}

The {Property \ref{eig}} shows that we can calculate the eigen-tensor from the flattened tensor to simplify the decomposition operations. Moreover, the {M-JGFT} is the bijection of GFT in the flattened MLG, with vertices indexed by both the layers and the entities. However, such {M-JGFT} analyzes the inter- and intra- layer connections jointly while ignoring the individual features of entities and layers. Next, we will show how to implement the order-wise frequency analysis in M-GSP based on tensor decomposition.

\subsection{Order-wise Spectral Analysis in M-GSP}\label{OAA}

In an undirected multilayer graph, the
representing tensor (adjacency/Laplacian) $\mathbf{F}$ is partially
symmetric between orders one and three, and 
between orders two and four, respectively. Then, the representing tensor can be written with the consideration of the multilayer graph structure under orthogonal CP-decomposition \cite{c28} as follows:
\begin{align}\label{cp_dec}
\mathbf{F}&\approx\sum_{\alpha=1}^{M}\sum_{i=1}^N \lambda_{\alpha i} \cdot \mathbf{f}_\alpha \circ\mathbf{e}_i\circ \mathbf{f}_\alpha \circ\mathbf{e}_i\\
&=\sum_{\alpha=1}^M\sum_{i=1}^N \lambda_{\alpha i} \tilde{\mathbf{V}}_{\alpha i} \circ\tilde{\mathbf{V}}_{\alpha i},\label{decompose1}
\end{align}
where $\mathbf{f}_\alpha \in\mathbb{R}^M$ are orthonormal, $\mathbf{e}_i\in\mathbb{R}^N$ are orthonormal and $\tilde{\mathbf{V}}_{\alpha i}=\mathbf{f}_{\alpha}\circ \mathbf{e}_i\in\mathbb{R}^{M\times N}$.

The CP decomposition factorizes a tensor into a sum of component rank-one
tensors, which describe the underlying features of each order. Although approximated algorithms are implemented to obtain the optimal decomposition, CP decomposition still achieves great success in real scenarios, such as feature extraction \cite{d5} and tensor-based PCA analysis \cite{d4}. A detailed discussion
of tensor decomposition and its implementation in M-GSP are provided in {Section \ref{td_compare}}.
In Eq. (\ref{decompose1}), $\mathbf{f}_\alpha$ and $\mathbf{e}_i$ capture the features of layers and entities, respectively, which can be interpreted as the subspaces of the MLG. More discussions about the frequency interpretation of order-wise M-GSP spectrum and connections to MWGSP spectrum are presented in Section \ref{dep}.

Note that, if there is only one layer in the multilayer graph, Eq. (\ref{decompose1}) reduces to the {eigen-decomposition} of a normal 
single-layer graph, i.e.,
$\mathbf{F}=\sum_{i=1}^N\lambda_i\mathbf{e}_i \circ \mathbf{e}_i$

With the decomposed representing tensor in Eq. (\ref{decompose1}), the order-wise MLG spectrum is defined as follows. 
\begin{definition}[Order-wise MLG Spectral Pair]
	For a multilayer graph $\mathcal{M}=\{\mathcal{V},\mathcal{L},\mathbf{F}\}$ with $M$ layers and $N$ nodes, the order-wise MLG spectral pairs are defined by $\{\lambda_{\alpha i},\mathbf{f}_\alpha,\mathbf{e}_i\}$,
	where $\{\mathbf{f}_1,\cdots,\mathbf{f}_M\}$ and $\{\mathbf{e}_1,\cdots,\mathbf{e}_N\}$ characterize features of layers and entities, respectively.
\end{definition}

With the definition of order-wise MLG spectral pair, we now explore their properties. Considering $\tilde{\mathbf{V}}_{\alpha i}=\mathbf{f}_{\alpha}\circ \mathbf{e}_i$, we have the following property, which indicates the availability of a joint MLG analysis based on order-wise spectrum.
\begin{property}
	The factor tensor $\tilde{\mathbf{V}}_{\alpha i}$ of the representing tensor $\mathbf{F}$ is the approximated eigen-tensor of $\mathbf{F}$.
\end{property}
\begin{proof}
Suppose that $\tilde{\mathbf{V}}_{\alpha i}$ is one factor tensor of $\mathbf{F}$ obtained from Eq. (\ref{decompose1}).

Let $\delta[k]$ denote the Kronecker delta. 
Since $\mathbf{f}_\alpha$ forms an orthonormal basis, then
the inner product would satisfy
\[<\mathbf{f}_\alpha, \mathbf{f}_{\beta}>=\sum_k [f_\alpha]_k\cdot[f_\beta]_k=\delta[\alpha-\beta].
\]
Similarly, 
\[ <\mathbf{e}_i, \mathbf{e}_j>=\delta [i-j].
\]
\begin{align}
[\mathbf{F}\diamond\tilde{\mathbf{V}}_{\alpha i}]_{\beta j}=&\sum_{\sigma=1}^M\sum_{ k=1}^{N} F_{\beta j \sigma k}\;[\tilde{V}_{\alpha i}]_{\sigma k}\\
\approx&\sum_{\sigma=1}^M\sum_{k=1}^{N} \sum_{\gamma=1}^M\sum_{t=1}^{N}\lambda_{\gamma t}\cdot\\
&[f_{\gamma}]_\beta\;[e_t]_{j}\;[f_{\gamma}]_\sigma\;[e_t]_{k}\;[\tilde{V}_{\alpha i}]_{\sigma k}.
\end{align}
Then, we have
\begin{align}\label{term}
\sum_{\sigma=1}^M\sum_{k=1}^{N}[f_{\gamma}]_\sigma\;[e_t]_{k}\;[\tilde{V}_{\alpha i}]_{\sigma k}
&=\sum_{\sigma=1}^M\sum_{k=1}^{N}[f_{\gamma}]_\sigma\;[e_t]_{k}\;[f_{\alpha}]_\sigma\;[e_i]_{k}\nonumber\\
&=\sum_{\sigma=1}^{M}[f_{\gamma}]_\sigma\;[f_{\alpha}]_\sigma\; \sum_{k=1}^{N}[e_t]_{k}\;[e_i]_{k}\nonumber\\
&=<\mathbf{f}_\gamma,\mathbf{f}_\alpha>\cdot <\mathbf{e}_t,\mathbf{e}_i>
\nonumber\\
&=\delta[\gamma-\alpha]\delta[t-i]
\end{align}
Thus,
\begin{equation}
[\mathbf{F}\diamond\tilde{\mathbf{V}}_{\alpha i}]_{\beta j}\approx \lambda_{\alpha i}[f_\alpha]_{\beta}[e_i]_j,
\end{equation}
which indicates
\begin{equation}
\mathbf{F}\diamond\tilde{\mathbf{V}}_{\alpha i}\approx\lambda_{\alpha i}\tilde{\mathbf{V}}_{\alpha i}.
\end{equation}
Then, $\tilde{\mathbf{V}}_{\alpha i}$ is the approximated eigen-tensor.
\end{proof}

This property indicates the relationship between the order-wise MLG spectral pair and the joint eigen-tensors.

By constructing a {fourth-order} tensor $\tilde{\mathbf{U}}_{\mathcal{F}}\in\mathbb{R}^{M\times N\times 
	M \times N}$ with $\tilde{\mathbf{V}}_{\alpha i}$ as its elements, i.e.,
$[\tilde{U}_{\mathcal{F}}]_{\alpha i \beta j}=[\tilde{V}_{\alpha i}]_{\beta j}$,
we can have the following property.

\begin{property}\label{ortho}
	Let $\mathbf{W}=\tilde{\mathbf{U}}_{\mathcal{F}}\otimes \tilde{\mathbf{U}}_{\mathcal{F}}$, where $\otimes$ is the contraction in the third and forth order with $W_{\alpha i \beta j}=\sum_{p,\theta}[\tilde{U}_{\mathcal{F}}]_{\beta j \theta p}\times [\tilde{U}_{\mathcal{F}}]_{\alpha i \theta p}$. Then, $\mathbf{W}$ is super-diagonal with super-diagonal elements 
	all equal to one.
\end{property}
\begin{proof}
	Let $\tilde{\mathbf{V}}_k=\tilde{\mathbf{V}}_{\alpha i}=\mathbf{f}_\alpha\circ \mathbf{e}_i$ and $\tilde{\mathbf{V}}_t=\tilde{\mathbf{V}}_{\beta j}=\mathbf{f}_\beta \circ \mathbf{e}_j$. Then, we have
	\begin{align}
	W_{\alpha i \beta j}&=\sum_{\theta, p}[\tilde{V}_k]_{\theta p}[\tilde{V}_t]_{\theta p}	\nonumber\\
	&=\sum_p [e_i]_p[e_j]_p\sum_\theta[f_\alpha]_\theta[f_\beta]_\theta
	\nonumber \\
	&=<\mathbf{f}_\alpha,\mathbf{f}_\beta>\cdot<\mathbf{e}_i,\mathbf{e}_j>,
	\nonumber\\
	&= \delta[\alpha-\beta]\delta[i-j].
	\end{align}
\end{proof}
This property generalizes the orthogonality of the spectral tensor into a similar
definition of matrix.

We now introduce the order-wise MLG spectral transform. Similar to Eq. (\ref{M-GFT}), the joint transform can be defined as
\begin{equation}\label{O-GFT}
\hat{\mathbf{s}}=\tilde{\mathbf{U}}_\mathcal{F} \diamond \mathbf{s}.
\end{equation}
Note that each element of $\hat{\mathbf{s}}$ in Eq. (\ref{O-GFT})
can be calculated as
\begin{align}
\hat{s}_{\alpha i}&=\sum_{\beta,j} [\tilde{U}_{\mathcal{F}}]_{\alpha i \beta j}s_{\beta j}\\
&=\sum_{\beta,j} [\tilde{V}_{\alpha i}]_{\beta j}s_{\beta j}\\
&=\sum_{\beta,j} [f_\alpha]_\beta \cdot [e_i]_j\cdot s_{\beta j}.
\end{align}
Let $\mathbf{E}_f=[\mathbf{f}_1\cdots\mathbf{f}_M]\in\mathbb{R}^{M\times M}$ and $\mathbf{E}_e=[\mathbf{e}_1\cdots\mathbf{e}_N]\in\mathbb{R}^{N\times N}$. We then have
\begin{equation}
	\hat{\mathbf{s}}'=\mathbf{E}_f^{\mathrm{T}}\mathbf{s}\mathbf{E}_e,
\end{equation}
with each element
\begin{equation}
	\hat{s}_\alpha'=\sum_{j,\beta}[f_\alpha]_\beta \cdot [e_i]_j\cdot s_{\beta j}
\end{equation}
Clearly, the M-GFT can be obtained as
\begin{equation}
	\hat{\mathbf{s}}=\hat{\mathbf{s}}'=\mathbf{E}_f^{\mathrm{T}}\mathbf{s}\mathbf{E}_e.
\end{equation}

Then, we have the following definition of M-GFT based on order-wise spectrum.

\begin{definition} [Order-wise M-GFT]
	Given the spectral vectors $\mathbf{E}_f=[\mathbf{f}_1\cdots\mathbf{f}_M]\in\mathbb{R}^{M\times
		M}$ and $\mathbf{E}_e=[\mathbf{e}_1\cdots\mathbf{e}_N]\in\mathbb{R}^{N\times N}$, the layer-wise M-GFT can be defined as
	\begin{equation}
	\hat{\mathbf{s}}_L=\mathbf{E}_f^{\mathrm{T}}\mathbf{s}\in\mathbb{R}^{M\times N},
	\end{equation}
	and the entity-wise M-GFT can be defined as 
	\begin{equation}
	\hat{\mathbf{s}}_N=\mathbf{s}\mathbf{E}_e\in\mathbb{R}^{M\times N}.
	\end{equation}
	The general M-GFT based on order-wise spectrum is defined as 
	\begin{equation} \label{ffff}
	\hat{\mathbf{s}}=\mathbf{E}_f^{\mathrm{T}}\mathbf{s}\mathbf{E}_e\in\mathbb{R}^{M\times N}.
	\end{equation}
\end{definition}
If there is only one layer in the multilayer graph, the M-GFT calculated with $\mathbf{s}^T\in\mathbb{R}^{N}$ as
$(\hat{\mathbf{s}}_N)^{\mathrm{T}}=(\mathbf{s}\mathbf{E}_e)^{\mathrm{T}}\in\mathbb{R}^{N}$, which
has the same form as the traditional GFT in Eq. (\ref{GFT}).

In addition, since $\mathbf{f}_\alpha$ and $\mathbf{e}_i$ are orthonormal basis of undirected MLG, the inverse M-GFT can be calculated as 
\begin{equation}\label{igft}
\mathbf{s}'=\mathbf{E}_f\hat{\mathbf{s}}\mathbf{E}_e^{\mathrm{T}}.
\end{equation}
Different from joint MLG Fourier space in Section \ref{JSA}, the order-wise MLG spectrum provides an individual analysis on layers and entities separately, and a reliable approximated analysis on the underlying MLG structures jointly. To minimize confusion, we abbreviate joint M-GFT in Eq. (\ref{M-GFT}) as M-JGFT. M-GFT refers to the order-wise transform in Eq. (\ref{ffff}) in the remaining parts if there is no specification.

\subsection{MLG Singular Tensor Analysis}
In addition to the eigen-decomposition, the singular value decomposition (SVD) is another important decomposition to factorize a matrix. In this part, we provide the higher-order SVD (HOSVD) \cite{c27} of the representing tensor as an alternative definition of spectrum for the multilayer graphs. 

Given the multilayer graph $\mathcal{M}=\{\mathcal{V},\mathcal{L},\mathbf{F}\}$ with $M$ layers and $N$ nodes in each layer, its representing tensor $\mathbf{F}\in\mathbb{R}^{M
	\times N\times M \times N}$ 
can be decomposed via HOSVD as
\begin{equation}\label{decomposeS}
\mathbf{F}= \mathbf{S}\times_1 \mathbf{U}^{(1)}\times_2 \mathbf{U}^{(2)}\times_3 \mathbf{U}^{(3)}\times_4 \mathbf{U}^{(4)},
\end{equation}
where $\mathbf{U}^{(n)}=[\mathbf{U}^{(n)}_1\quad \mathbf{U}^{(n)}_2\quad\cdots\quad \mathbf{U}^{(n)}_{I_n}]$ is a unitary $(I_n\times I_n)$ matrix, with $I_1=I_3=M$ and $I_2=I_4=N$.
$\mathbf{S}$ is a complex $(I_1\times I_2\times I_3 \times I_4)$-tensor of which the subtensor $\mathbf{S}_{i_n}$ obtained by fixing $n$th index to $\alpha$ have
\begin{itemize}
	\item $<\mathbf{S}_{i_n=\alpha},\mathbf{S}_{i_n=\beta}>=0$ where $\alpha\neq\beta$.
	\item $||\mathbf{S}_{i_n=1}||\geq||\mathbf{S}_{i_n=2}||\geq\cdots\geq ||\mathbf{S}_{i_n=I_n}||\geq 0$.
\end{itemize}

The Frobenius-norms $\sigma_i^{(n)}=||\mathbf{S}_{i_n=i}||$ is the $n$-mode singular value, and $\mathbf{U}^{(i)}$ are the corresponding $n$-mode singular vectors. For an undirected multilayer graph, the representing tensor is symmetric for
every 2-D combination. Thus, there are two modes of singular spectrum, i.e., $(\gamma_\alpha, \mathbf{f}_\alpha)$ for mode $1,3$, and $(\sigma_i,\mathbf{e}_i)$ for mode $2,4$. 
More specifically, $\mathbf{U}^{(1)}=\mathbf{U}^{(3)}=(\mathbf{f}_\alpha)$ and $\mathbf{U}^{(2)}=\mathbf{U}^{(4)}=(\mathbf{e}_i)$.
Since the joint singular tensor captures the consistent information of entities and layers, it can be calculated as 
\begin{equation}\label{decom}
(\lambda_{\alpha i}, \hat{\mathbf{V}}_{\alpha i})=(\gamma_\alpha\cdot\sigma_i, \mathbf{f}_\alpha \circ \mathbf{e}_i).
\end{equation}
Note that the diagonal entries
of $\mathbf{S}$ are not the eigenvalues or frequency coefficients of the representing tensor in general. The multilayer graph singular space is defined as follows.
\begin{definition}[Multilayer Graph Singular Space]
	For a multilayer graph $\mathcal{M}=\{\mathcal{V},\mathcal{L},\mathbf{F}\}$ with $M$ layers and $N$ nodes, the MLG singular space is defined as the space consisting of all singular tensors $\{\hat{\mathbf{V}}_1\cdots\hat{\mathbf{V}}_{MN}\}$ obtained from Eq. (\ref{decom}). The singular vectors $\{\mathbf{f}_1,\cdots,\mathbf{f}_M\}$ and $\{\mathbf{e}_1,\cdots,\mathbf{e}_N\}$ in Eq. (\ref{decomposeS}) characterize layers and entities, respectively.
\end{definition}

Similar to order-wise spectral analysis in Section \ref{OAA},
we can define the MLG singular tensor transform (M-GST) based on the singular tensors as follows.
\begin{definition}[M-GST]
	Suppose that $\mathbf{U}_s=(\mathbf{f}_\alpha\circ\mathbf{e}_i)
	\in\mathbb{R}^{ M\times N \times M \times N}$ consists of the singular vectors of the representing tensor $\mathbf{F}$ in Eq. (\ref{decomposeS}), where	
	$[U_s]_{\alpha i \beta j}=[f_\alpha]_\beta\cdot [e_i]_j$.	
	The M-GST can be defined as the contraction between $\mathbf{U}_s$ and the tensor signal $\mathbf{s}\in\mathbb{R}^{M\times N}$, i.e.,
	\begin{equation} \label{ssss}
	\check{\mathbf{s}}=\mathbf{U}_s \diamond \mathbf{s}.
	\end{equation}
	If the singular vectors are included in $\mathbf{W}_f=[\mathbf{f}_1\cdots\mathbf{f}_M]\in\mathbb{R}^{M\times 
		M}$ and $\mathbf{W}_e=[\mathbf{e}_1\cdots\mathbf{e}_N]\in\mathbb{R}^{N\times N}$, the layer-wise M-GST can be defined as
	\begin{equation}
	\check{\mathbf{s}}_L=\mathbf{W}_f^{\mathrm{T}}\mathbf{s}\in\mathbb{R}^{M\times N},
	\end{equation}
	and the entity-wise M-GST can be defined as 
	\begin{equation}
	\check{\mathbf{s}}_N=\mathbf{s}\mathbf{W}_e\in\mathbb{R}^{M\times N}.
	\end{equation}
\end{definition}
\noindent
Inverse M-GST can be defined similarly as in Eq. (\ref{igft}) with unitary $\mathbf{W}_e$ and $\mathbf{W}_f$.

Compared to the eigen-tensors in Eq. (\ref{ted}), the singular tensors come from the combinations of the singular vectors, thus are capable 
of capturing information of layers and entities more efficiently. Eigen-decomposition, however, focuses more on the joint information and approximate the separate information of layers and entities.
We shall provide further discussion on 
the performance of different decomposition methods in {Section \ref{td_compare}}. The intuition of applying HOSVD in MLG analysis and its correlations to GSP are also provided in Section \ref{i_hosvd}.

\subsection{Spectrum Ranking in the Multilayer Graph} \label{ex_fre}
In traditional GSP, the frequencies are defined by the eigenvalues of
the shift, whereas
the total variation is an alternative measurement 
of the order of the graph frequencies \cite{c2}. 
Similarly, we use the total variation of $\lambda_{\alpha i}$ based on the spectral tensors to rank the MLG frequencies. Let $|\lambda|_{max}$ be the joint eigenvalue {of the adjacency tensor $\mathbf{A}$} with the largest magnitude.
The M-GSP total variation is defined as follows:
\begin{align}\label{tv}
TV(\mathbf{V}_{\alpha i})&=||\mathbf{V}_{\alpha i}-\frac{1}{|\lambda|_{max}} \mathbf{A}\diamond \mathbf{V}_{\alpha i}||_1\\
&=|1-\frac{\lambda}{|\lambda|_{max}}\mathbf|\cdot||\mathbf{V}_{\alpha i}||_1,
\end{align}
where $||\cdot||_1$ is the {element-wise} $l_1$ norm. Other norms could also be used to define the total variation.
For example, the $l_2$ norm could be efficient in signal denoising \cite{c2}. The graph frequency related to $\lambda_{\alpha i}$ is said to be a higher frequency if its total variation $TV(\mathbf{V}_{\alpha i})$ is larger, and its corresponding spectral tensor $\mathbf{V}_{\alpha i}$ is a higher frequency spectrum. 
{If the representation tensor refers to Laplacian tensor, i.e., $\mathbf{L=D-A}$, the frequency order is in contract to the adjacency tensor $\mathbf{A}$ as GSP \cite{c2}}.
We 
shall provide more details on 
interpretation of MLG frequency in Section \ref{fre_int}.

\section{Filter Design} \label{fter}
In this section, we introduce an MLG filter design together with its properties based on signal shifting.

\subsection{Polynomial Filter Design}
Polynomial filters are  basic filters in 
GSP \cite{c7,d6}. In M-GSP, 
first-order filtering consists of
basic signal filtering, i.e.,
\begin{equation}
    \mathbf{s}'=f_1(\mathbf{s})=\mathbf{F}\diamond \mathbf{s}.
\end{equation}
Similarly, a second order filter can be defined as 
additional filtering on first-order filtered signal, i.e.,
\begin{align}
\mathbf{s}''&=f_2(\mathbf{s})\\
&=\mathbf{F}\diamond(\mathbf{F}\diamond \mathbf{s}),
\end{align}
whose entries $s_{\alpha i}''$ are calculated as
\begin{align}
s_{\alpha i}''&=\sum_{\beta=1}^{M}\sum_{j=1}^N F_{\alpha i \beta j} s_{\beta j}'\\
&=\sum_{\beta ,j} F_{\alpha i \beta j}\sum_{\epsilon ,p} F_{\beta j \epsilon p}s_{\epsilon p}\\
&=\sum_{\epsilon, p}s_{\epsilon p}\sum_{\beta, j} F_{\alpha i \beta j} F_{\beta j \epsilon p}\\
&=(\mathbf{F}\odot\mathbf{F})\diamond \mathbf{s},
\end{align}
where $\odot$ is the contraction defined in Eq. (\ref{oodot}).

Let $\mathbf{F}^{[2]}=\mathbf{F}\odot \mathbf{F}$. From Eq. (\ref{ted}),
we have:
\begin{align}
F^{[2]}_{\alpha i\beta j}&=\sum_{\theta,p} F_{\alpha i \theta p} F_{\theta p \beta j}\\
&=\sum_{\theta,p}(\sum_k\lambda_k[{V}_k]_{\alpha i}[V_k]_{\theta p})(\sum_t \lambda_t[V_t]_{\beta j}[V_t]_{\theta p})\nonumber\\
&=\sum_{k,t}\lambda_k\lambda_t[ {V}_k]_{\alpha i}[ {V}_t]_{\beta j}(\sum_{\theta,p}[{V}_t]_{\theta p}[{V}_k]_{\theta p})\nonumber\\
&=\sum_k \lambda_k^2[ {V}_k]_{\alpha i}[ {V}_k]_{\beta j}.
\end{align}
Similarly, for $\tau$th-order term $\mathbf{F}^{[\tau]}$, its entry $F^{[\tau]}_{\alpha i\beta j}$ can be calculated as
$F^{[\tau]}_{\alpha i\beta j}=\sum_k \lambda_k^\tau[\mathbf{V}_k]_{\alpha i}[\mathbf{V}_k]_{\beta j}$.

Now we have the following property.
\begin{property}
	The $\tau$-th order basic shifting filter $f_\tau(\mathbf{s})$ can be calculated as
	\begin{align}\label{polyf}
	f_\tau(\mathbf{s})&=\mathbf{F}^{[\tau]}\diamond\mathbf{s}\\
	&=(\sum_{k=1}^{MN}\lambda_k^{\tau}\mathbf{V}_k\circ\mathbf{V}_k)\diamond\mathbf{s}.
	\end{align}
\end{property}
This property is the M-GSP counterpart to the traditional
linear system interpretation
that complex
exponential signals are eigenfunctions of linear systems \cite{c2}, and provide a quicker implementation of higher-order shifting.
With the $k$-order polynomial term, the adaptive polynomial filter is defined as 
\begin{equation}\label{adapt}
h(\mathbf{s})=\sum_k \alpha_k\mathbf{F}^{[k]}\diamond\mathbf{s},
\end{equation}
where $\{\alpha_k\}$ are parameters to be estimated from data.

Adaptive polynomial filter is useful in semi-supervised classification \cite{d7} and exploits
underlying geometric topologies. We will illustrate further and provide application examples based on MLG polynomial filtering in Section \ref{app}.

\subsection{Spectral Filter Design}\label{fefi}
Filtering in the graph spectrum space is useful in GSP frequency analysis. For example, ordering the Laplacian graph spectrum $\mathbf{V}_\mathcal{G}=[\mathbf{e}_1, \cdots, \mathbf{e}_N]\in\mathbb{R}^{N\times N}$ in a descent order by the graph total variation \cite{c2}, i.e., high frequency to low frequency, the GFT of $\mathbf{s}\in\mathbb{R}^{N}$ is calculated as $\hat{\mathbf{s}}=\mathbf{V}_\mathcal{G}^\mathrm{T}\mathbf{s}$. By removing $k$ elements in the low frequency part, i.e., $\hat{s}'=[\hat{s}_1,\cdots, \hat{s}_{N-k}, 0, \cdots, 0]$, a high-pass filter can be designed as
\begin{align}
\mathbf{s}'&=\mathbf{V}_\mathcal{G}\hat{s}'\\
&=\mathbf{V}_\mathcal{G}\Sigma_k\mathbf{V}_\mathcal{G}^\mathrm{T}\mathbf{s}
\end{align}
where $\Sigma_k=diag([\sigma_1,\cdots,\sigma_N])$ is a diagonal matrix with $\sigma_i=0$ for $i=1,\cdots,N-k$; otherwise, $\sigma_i=0$.

Similarly, in M-GSP, a spectral filter is designed by filtering in the spectrum space together with the inverse M-GFT. With Eq. (\ref{ffff}) and Eq. (\ref{ssss}),  spectral filtering of $\mathbf{s}$ is defined as
\begin{align}
&\mathbf{s}'=\nonumber\\
&\mathbf{E}_f
\begin{bmatrix}
g(\gamma_1) & \cdots & 0\\
\vdots & \ddots & \vdots\\
0&\cdots&g(\gamma_M)
\end{bmatrix}	
\mathbf{E}_f^{\mathrm{T}}\mathbf{s}
\mathbf{E}_e
\begin{bmatrix}
f(\sigma_1) & \cdots & 0\\
\vdots & \ddots & \vdots\\
0&\cdots&f(\sigma_N)
\end{bmatrix}	\mathbf{E}_e^{\mathrm{T}}\label{mgsp_sf}
\end{align}
where functions $g(\cdot)$ and $f(\cdot)$ are designed by the specific tasks.
For example, if one wants to design a layer-wise filter capturing the smoothness of signals in the MLG singular space, the function $g(\cdot)$ can be designed as $\mathbf{g}=[1,\cdots,1, 0, \cdots, 0]$ by ordering the layer-wise singular vectors in the descent order of singular values. 

\subsection{Discussion}
{We briefly discuss
the interpretation of polynomial filters and spectral filters.
From the spatial perspective, MLG polynomial filter is a weighted sum of messaging passing on the multilayer graph in different orders, shown as Eq.~(\ref{adapt}). Each node collects information from both inter- and intra- layer neighbors, before combining them with its own information. From the spectrum 
perspective, M-GSP polynomial filters are eigenfunctions of linear systems, which are special cases of M-GSP spectral filters shown in Eq. (\ref{mgsp_sf}) The M-GSP spectral filters assign different weights to each M-GSP spectrum via functions ${f(\cdot)}$ and ${g(\cdot)}$ depending on specific tasks. Both M-GSP polynomial filters and spectral filters can be useful for high-dimensional IoT signal processing. More discussions and examples of M-GSP filters are presented in Section \ref{app}.}

\section{Discussion and Interpretative Insights} \label{discus}
\subsection{Interpretation of M-GSP Frequency} \label{fre_int}

\subsubsection{Interpretation of Graph Frequency}\label{gfe}
To better understand its physical meaning, we start with the total variation in digital signal processing (DSP). The total variation in DSP is defined as
differences among  signals over time \cite{c34}. 
Moreover, the total variations of frequency components
have a 1-to-1 correspondence to frequencies in the order
of their values. If the total variation of a frequency component
is larger, the corresponding frequency with the same index
is higher. This means that, a higher
frequency component changes faster over time and exhibits
a larger total variation. Interested readers could refer to \cite{c2,c8} for a detailed interpretation of total variation in DSP.

\begin{figure}[t]
	\centering
	\includegraphics[width=1.5in]{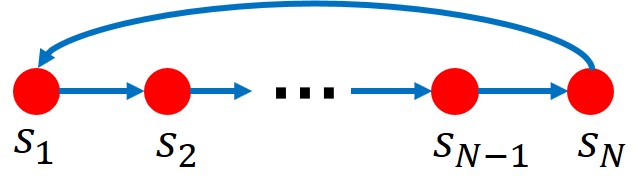}
	\caption{Example of Cyclic Graph.}
	\label{cirg}
\end{figure}

Now, let us elaborate the graph frequency motivated by the cyclic graph. Rewrite the finite signals in DSP as vectors, i.e., $\mathbf{s}=[s_1, \cdots,s_{N}]\in\mathbb{R}^{N}$, the signal shifting can be interpreted as the shift filtering corresponding to a cyclic graph shown in Fig. \ref{cirg}. Suppose that its adjacency matrix is written as
\begin{align}
{
	\mathbf{C}_N=
	\begin{bmatrix}
	0&0&\cdots&0&1\\
	1&0&\cdots&0&0\\
	\vdots&\ddots&\ddots&\ddots&\vdots\\
	0&0&\ddots&0&0\\
	0&0&\cdots&1&0
	\end{bmatrix}}
\end{align}
Then, the shifted signal over the cyclic graph is calculated as $\mathbf{s}'=\mathbf{C}_N\mathbf{s}=[s_{N}\quad s_1\quad\cdots\quad s_{N-1}]^{\mathrm{T}}$, which shifts the signal at each node to its next node. More specifically,
$\mathbf{C}_N$ can be decomposed as $\mathbf{C}_N=\mathbf{V\Lambda}\mathbf{V}^{-1}$ where the eigenvalues $\lambda_n=e^{-j\frac{2\pi n}{N}}$ and $\mathbf{V}^{-1}=\frac{1}{\sqrt{N}}[\lambda^{kn}_N]$ is the discrete Fourier matrix. 
Inspired by the DSP, the eigenvectors in $\mathbf{V}$ are the spectral components (spectrum) of the cyclic graph and the eigenvalues are related to the graph frequencies \cite{c2}.

Generalizing the adjacency matrix of the cyclic graph to the representing matrix $\mathbf{F}_M$ of an arbitrary graph, the graph Fourier space consists of the eigenvectors of $\mathbf{F}_M$ and the graph frequencies are related to the eigenvalues. 
More specifically, the graph Fourier space can be interpreted as the manifold or spectrum space of the representing matrix.
As aforementioned, the total variations of frequency components
reflect the order of frequencies, we can also use the total variation, i.e.,
$TV(\mathbf{e}_i)=||\mathbf{e}_i-\frac{1}{|\lambda|_{max}}\mathbf{e}_i||_1$,
to rank the graph frequencies, where $\mathbf{e}_i$ is the spectral component related to the eigenvalue $\lambda_i$ in $\mathbf{F}_M$.
Similar to DSP, the graph frequency indicates the oscillations over the vertex set, i.e., how fast the signals change over the graph shifting.

\subsubsection{Interpretation of MLG Frequency}
Now, return to M-GSP. Given spectral tensors $\mathbf{V}_k\in\mathbb{R}^{M\times N}$ of a multilayer graph, a signal $\mathbf{s}\in\mathbb{R}^{M\times N}$ can be written in a weighted sum of the spectrum, i.e.,
$\mathbf{s}=\sum_k a_k\mathbf{V}_k$.
Viewing the spectral tensor as a signal component, the total variation is in the form of differences between the original signal and its shifted version in Eq. (\ref{tv}). If the signal component changes faster over the multilayer graph, the corresponding total variation is larger. Since we 
relate
higher frequency component with a larger total variation, the MLG frequency indicates how fast the signal propagates over the multilayer graph under the representing tensor. If a signal $\mathbf{s}$ contains more high frequency components, 
it changes faster under the representing tensor.

\subsubsection{Interpretation of MLG Singular Tensors}\label{i_hosvd}
As discussed in Section \ref{gfe}, the name of graph Fourier space arises from the adjacency matrix of the cyclic graph. However, when the algebra representation is generalized to an arbitrary graph, especially the Laplacian matrix, the definition of graph spectrum is less related to the Fourier space in DSP but can be interpreted as the manifold or subspace of the representing matrix instead. In literature, SVD is an efficient method to obtain the spectrum for signal analysis, such as spectral clustering \cite{d11} and PCA analysis \cite{d12}. It is straightforward to generalize graph spectral analysis to graph singular space, especially for the Laplacian matrix. In M-GSP, the order-wise singular vectors can be interpreted as subspaces characterizing features of layers and entities, respectively. Since HOSVD is robust and efficient, transforming signals to the MLG singular space (M-GST) for the analysis of underlying structures can be a useful alternative for M-GFT.

\subsection{Interpretation of Entities and Layers} \label{exn_la}
\begin{figure}[t]
	\centering
	\includegraphics[width=2in]{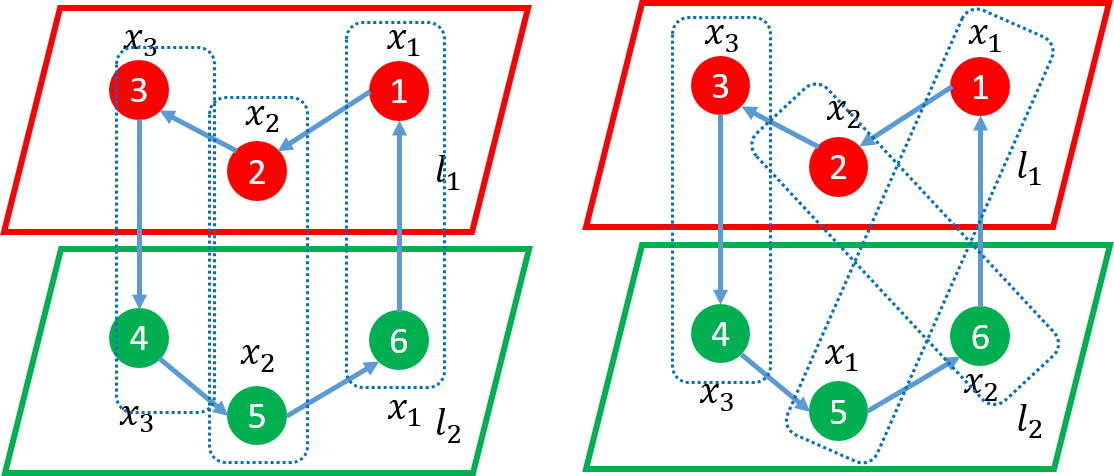}
	\caption{Example of Different Entities.}
	\label{le}
\end{figure}

To gain better physical insight of entities and layers, 
we discuss two categories of datasets:
\begin{itemize}
	\item In most of the physical systems and datasets, 
	signals can be modeled with a specific physical meaning in terms
	of layers and entities. In smart grid, for example, each station can be an entity, connected in two layers of computation and
	power transmission, respectively. Another example is video in which each geometric
	pixel point is an entity and 
	each video frame form a layer.
	Each layer node denotes the pixel value in
	that video frame. M-GSP can be intuitive tool for these datasets and systems.
	
	\item In some scenarios, however, the datasets usually only has a definition of layers without meaningful entities. In particular, for  multilayer graphs with different numbers of nodes, 
	we may insert some isolated artificial 
	nodes to augment the multilayer graph.
	Often in such applications, it may be harder to identify the physical meaning of entities. Here, the entities may be virtual and 
	are embedded in the underlying structure of the multilayer graph. 
	Although definition of a virtual entity may vary with the chosen 
	adjacency tensor, it relates to the topological structure in terms
	of global spectral information. For example, in Fig. \ref{le}, 
	we can use two different definitions of virtual entities. Although the representing tensors for these two definitions differ, their eigenvalues 
	remain the same. Considering also layer-wise flattening, 
	the two supra-matrices
	are related by reshaping, by exchanging the fourth and fifth columns and rows.
	They still have the same eigenvalues, whereas the
	eigentensors can also be the same by implementing the reshaping operations.
	Note that, to capture distinct information 
	from entities, their spectra would change with
	different definitions of virtual entities.
\end{itemize}

\subsection{Distinctions from Existing GSP Works} \label{dep}
\subsubsection{Graph Signal Processing}
Generally, M-GSP extends traditional GSP into multilayer graphs. Although one can stack all MLG layers to represent them with a supra-matrix, such matrix representation 
makes GSP inefficient in extracting features of layers and entities separately. Given a supra-matrix of the MLG, the layers of nodes can not be identified directly from its index since all the nodes are treated equivalently. However, the tensor representation provides clear identifications on layers in its index. Moreover, in GSP, we can only implement a joint transform to process inter- and intra- layer connections together, while the M-GSP provide a more flexible choice on joint and order-wise analysis. In Section \ref{JSA}, the joint M-GSP analysis introduced can be viewed as the bijection of GFT in the flattened MLG, with vertices indexed by both layers and entities. Beyond that, we flexibly provide order-wise spectral analysis based on tensor decompositions, which allow the order-wise analysis on layers and nodes. One can select suitable MLG-based tools depending on tasks. The joint spectral analysis can be implemented if we aim to explore layers and entities fully, whereas the order-wise spectral and singular analysis are more efficient in characterizing layers and entities separately.

\subsubsection{Joint Time-Vertex Fourier Analysis}
In \cite{c19}, a joint time-vertex Fourier transform (JFT) is defined by implementing GFT and DFT consecutively.
As discussed in Section \ref{gfe}, the time sequences can be interpreted under a cyclic graph, and thus reside on a MLG structure. However, JFT assumes that all the layers have the same intra-layer connections, which limits the generalization of the MLG analysis. Differently, the tensor-based representation allows heterogeneous structures for the intra-layer connections, which makes M-GSP more general.

\subsubsection{Multi-way Graph Signal Processing}
In \cite{c18}, MWGSP 
has been proposed to process high-dimensional signals. Given $K$th-order high-dimensional signals, one can decompose the tensor signal in different orders, and construct one graph for each. Graph
signal is said to reside on a high-dimensional product graph obtained by the product of all individual factor graphs. 
Although the MW-GFT is similar to M-GFT for $K=2$, 
there still are notable differences in terms of spectrum. 
First, MWGSP can only process signals without exploiting a given structure 
since multiple graph spectra would arise from each order of the signals. 
For a multilayer graph with a given structure, such as physical networks with heterogeneous intralayer connections,
MWGSP does not naturally process it efficiently and cohesively. The order-wise spectra come from factor graphs of each order in MWGSP while M-GSP spectra are calculated from the tensor representation of the whole MLG.
Second, MWGSP assumes all the layers residing on a homogeneous factor graph and restricts the types of manageable
MLG structure. For example, in a spatial temporal dataset, 
a product graph, formed by the product of spatial connections and temporal connections, assumes the same topology in each layer. However, 
many practical systems and datasets feature more complex geometric interactions. M-GSP provide a more intuitive and natural 
framework for such MLG. 
In summary, despite some shared similarities
between MW-GFT and M-GFT in some scenarios, 
they serve different purposes and are suitable for different
underlying data structures.

\subsection{Comparison of Different Decomposition Methods} \label{td_compare}
{To compare recovery accuracy of representing tensor 
using different tensor decomposition methods, we
examine eigen-tensor decomposition, HOSVD, optimal CP decomposition and Tucker decomposition in MLGs randomly generated from the $Erd\ddot{o}s-R\dot{e}nyi$ (ER) random graph $ER(p,q,M,N)$. Here $M$ is the number of layers with $N$ nodes in each layer, 
$p$ is the intralayer connection probability and $q$ is the interlayer connection probability. 
We apply different decomposition methods of similar
complexity, and compute errors between
decomposed and original tensors.
From Table \ref{err}, 
we can see that the eigen-tensor decomposition and HOSVD 
exhibit better overall accuracy. Generally, 
eigen-tensor decomposition is better
suited for
applications emphasizing joint features of layers and entities.
On the other hand, HOSVD is 
effective at separating 
individual features of layers and entities.
Note that, in addition to recovery accuracy, different decompositions 
may have different performance 
advantages when capturing different data features
that can be better measured with different metrics. }

\begin{table}[t]
	\scriptsize
	\centering
	\caption{Error of Decomposing the Representing Tensor}
	\begin{tabular}{|l|l|l|l|}
		\hline
		Graph Structure           & ER(0.3,0.3,6,5) & ER(0.5,0.7,11,15) & ER(0.8,0.7,6,15) \\ \hline
		Eigen-tensor & 8.3893e-15    & 1.6001e-13      & 3.8347e-13     \\ \hline
		HOSVD                     & 1.011e-14    & 1.9563e-13      & 1.9056e-13     \\ \hline
		OPT-CP                    & 9.22e-01      & 8.82e-01        & 9.24e-01       \\ \hline
		Tucker                    & 9.37e-05      & 9.10e-05        & 9.40e-05       \\ \hline
	\end{tabular}
	\label{err}
\end{table}

\section{Application Examples} \label{app}
We now provide some illustrative application examples within our
M-GSP framework.

\subsection{Analysis of Cascading Failure in IoT Systems}
{
Analysis of cascading failure in IoT systems based on the spreading processes in multilayer graphs has recently attracted significant interests \cite{ca1}. Modeling the failure propagation in complex systems by shifting over multilayer graph, M-GSP spectrum theory can help the analysis of system stability.}
{
In this part, we introduce a M-GSP analysis for cascading failure over multilayer cyber-physical systems based on epidemic model \cite{c12}. Shown in Fig. \ref{ex2}, a cyber-physical system with $M$ layers and $N$ nodes in each layer can be intuitively modeled by a MLG with adjacency tensor $\mathbf{A}\in \mathbb{R}^{M\times N\times M \times N}$.}

{
Here, we consider the susceptible-infectious-susceptible (SIS) model \cite{keeling2011modeling} for the failure propagation. In the SIS model, each node has two possible states: susceptible (not fail) or infectious (fail). At each time slot, the infectious node may cause failure to other nodes through directed links at certain infection rates, or it may heal itself spontaneously at a self-cure rate. The initial attack make several nodes infectious. }

{
Since the nodes in the same layer correspond to the same functionality, e.g., power transmission, nodes on the same layer have the same self-cure rate and infection rate. The notations of the epidemic model are given as follows:
\begin{itemize}
	\item $\mu_{ \alpha}$:  self-cure rate for nodes on layer $ \alpha$;
	\item $\theta_{\alpha \beta}$:  infection rate describing failure propagation probability from nodes on layer $\beta$ to those on layer $\alpha$;
	\item $P_{i,\alpha}(t)$:  failure probability of the projected node of entity $i$ on layer $\alpha$ at time $t$;
	\item $\epsilon_{i,\alpha}(t)$:  transition probability that the projected node of entity $i$ on layer $\alpha$ shifts from infectious state to susceptible state at time $t$;
	\item $\sigma_{i,\alpha}(t)$:  transition probability that the projected node of entity $i$ on layer $\alpha$ remains susceptible at time $t$.
\end{itemize}}

{
Since an infectious node becomes susceptible if it cures itself without being infected by its neighbors,
we have
\begin{equation}
\epsilon_{i,\alpha}(t)=\mu_{ \alpha}\prod_{j,\beta}[1-\theta_{\alpha \beta}A_{\alpha i \beta j}P_{j,\beta}(t)]
\end{equation}
Similarly, a susceptible node remains susceptible without being infected by
its neighbors. Thus, 
\begin{equation}
\sigma_{i,\alpha}(t)=\prod_{j,\beta}[1-\theta_{\alpha \beta}A_{\alpha i \beta j}P_{j,\beta}(t)]
\end{equation}
The state transition forms a Markov chain, for which we derive the failure probability as
\begin{align}\label{trans}
P_{i,\alpha}(t)=1-\prod_{j,\beta}[1-T_{\alpha i \beta j}P_{j, \beta}(t-1)]\end{align}
where $T_{\alpha i \beta j}=(1-\mu_{\alpha})\delta_{\alpha i \beta j}+\theta_{\alpha \beta}A_{\alpha i \beta j}$,
with $\delta_{\alpha i \beta j}=1$ if $(j,\beta)=(i,\alpha)$; otherwise, $\delta_{\alpha i \beta j}=0$.}

{
We can define a transition tensor $\mathbf{T}\in \mathbb{R}^{M\times N\times M\times N}$ with elements $T^{\beta j}_{\alpha i}$ to characterize the failure propagation in Eq. (\ref{trans}). In steady state, $P_{i,\alpha}(\tau)=P_{i,\alpha}(\tau-1)$. Moreover,
\begin{equation}
\tilde P_{i,\alpha}=1-\prod_{j,\beta}[1-T^{\beta j}_{\alpha i}\tilde P_{j,\beta}],
\end{equation}
where $\tilde P_{i,\alpha}$ is the failure probability of the projected node of entity $i$ on layer $\alpha$ in steady state. 
Following \cite{c12}, we can arrive that $\tilde P_{i,\alpha}=0$ if the spectral radius of the transition tensor $\rho(\mathbf{T})<1$, which indicates no failed nodes in steady state. Thus, $\rho(\mathbf{T})$ could serve as an indicator for system robustness. Here,
to avoid being repetitive, we merely introduce a simple example of MLG-based cascading failure analysis. Interested readers may refer to our work \cite{c12} for more details. With a better understanding of M-GSP spatial shifting, one can develop more general analysis for various failure models in the multilayer IoT systems.
}

\begin{figure*}[htb]
	\centering
	\subfigure[Original Image.]{
		\label{CELL11}
		\includegraphics[height=6cm]{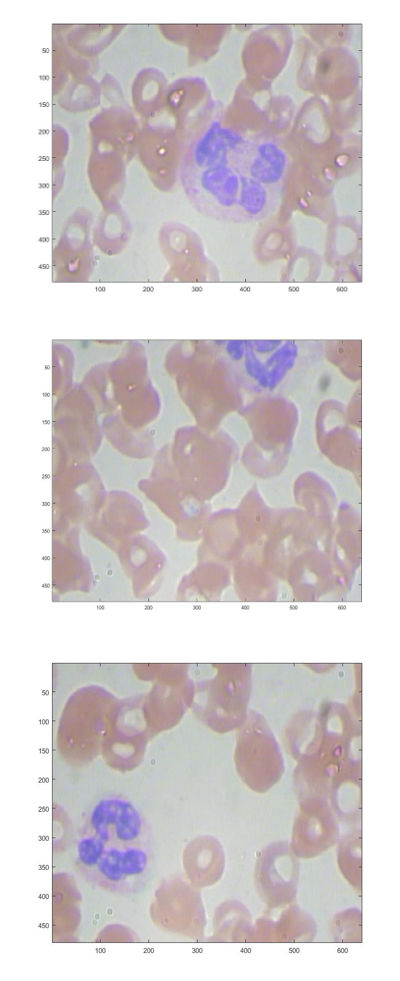}}
	\hfill
	\subfigure[K-Means.]{
		\label{CELL21}
		\includegraphics[height=6cm]{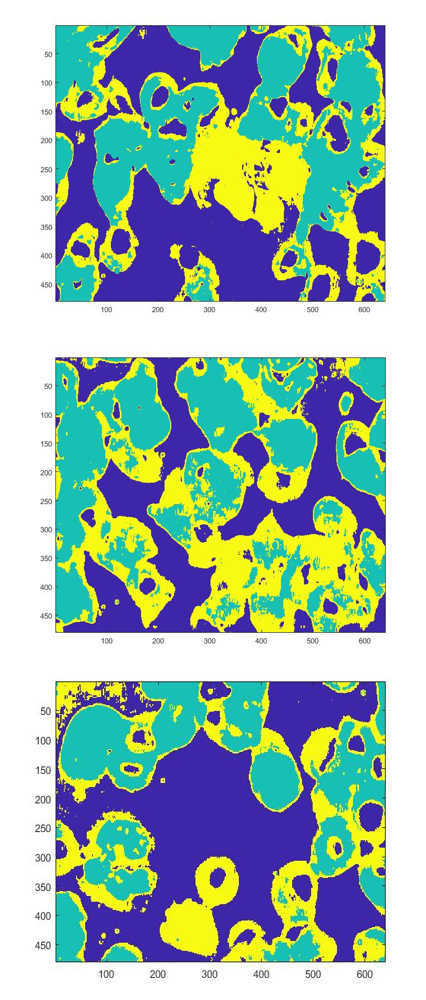}}
	\hfill
	\subfigure[GSP.]{
		\label{CELL31}
		\includegraphics[height=6cm]{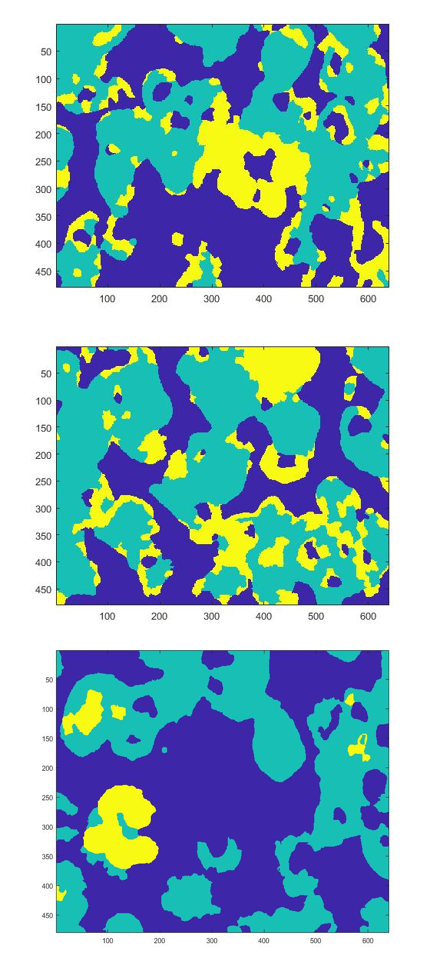}}
	\hfill
	\subfigure[MLG-SVD.]{
		\label{CELL41}
		\includegraphics[height=6cm]{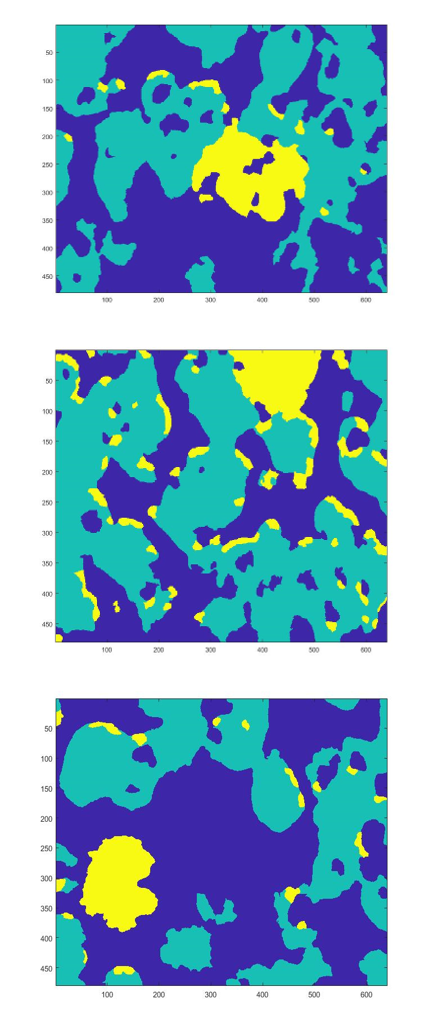}}
	\hfill
	\subfigure[MLG-FZ.]{
		\label{CELL5}
		\includegraphics[height=6cm]{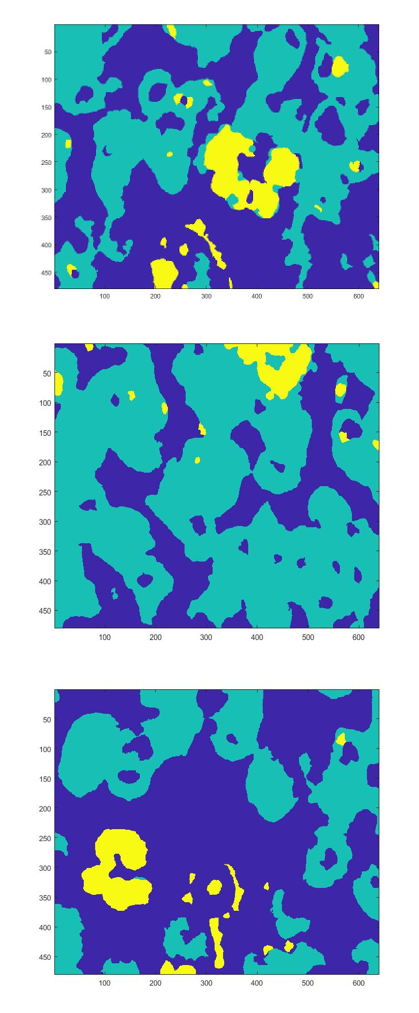}}
	\caption{Example of BCCD Datasets and Segmented Images: (a) the original image; (b)-(e) segmented images under different methods (WBCs are marked yellow, RBCs are marked green, and Platelet (P) is marked blue).}
	\label{CELL123}
\end{figure*}

\subsection{Spectral Clustering}
{Clustering is a widely used tool in a variety of applications such as social network analysis, computer vision,
and {IoT}. Spectral
clustering is popular and 
effective among many variants. Modeling
dataset by a normal graph before spectral clustering, significant performance improvement is possible in structured
data \cite{d11}. In this part, we introduce M-GSP spectral clustering and demonstrate its application in RGB image segmentation.}

To model an RGB image using MLG, we 
can directly treat its three colors as three layers. 
To reduce the number of nodes for computational efficiency, 
we first build $N$ superpixels for a given image and represent each superpixel as an entity in the multilayer graph. Here, we define the feature of a superpixel according to its RGB
pixel values. 
For interlayer connections, each node connects with its counterparts
in other layers. For intralayer connections on layer $\ell$, 
we calculate the Gaussian-based distance between two superpixels 
according to 
\begin{equation}
  W_{ij}=
\exp\left(-\frac{|\mathbf{s}_i(\ell)-\mathbf{s}_j(\ell)|^2}{\delta_\ell^2}\right)  
\end{equation}
if $|\mathbf{s}_i(\ell)-\mathbf{s}_j(\ell)|^2\leq t_\ell$; otherwise, $W_{ij}=0$,
where $\mathbf{s}_i(\ell)$ is the superpixel value on layer $\ell$,
$\delta_\ell$ is an adjustable parameter and $t_\ell$ is a predefined threshold. 
Different layers 
may have different structures. 
The threshold $t$ is set to be the mean of all 
pairwise distances. 
As such, an RGB image is modeled as multiplex graph
with $M=3$ and $N$ nodes. 

We now consider MLG-based spectral clustering. For image segmentation, we focus on the properties of entities (i.e., superpixels), and implement spectral clustering over
entity-wise spectrum by proposing Algorithm \ref{basic4}. The previous discussions
have been summarized in steps 1-3. In Step 4, different schemes may be used to calculate spectrum, including
spectral vector via tensor factorization in Eq. (\ref{decompose1}), and
singular vector in Eq. (\ref{decomposeS}).
Step 5 determines 
$K$ based on the largest arithmetic
gap in eigenvalues. 
Traditional clustering methods, such as $k$-means clustering \cite{d11}, can be carried out in Step 6.

\begin{table*}[t]
	\centering
	\scriptsize
	\caption{Results of Image Segmentation in Image BSD300 and BSD500}
	\begin{tabular}{|l|l|l|l|l|l|l|}
		\hline
		& BSD300(N=100, all)   & BSD300(N=300, all)   & BSD300(N=100, Coarse) & BSD300(N=300, Coarse) & BSD300(N=900,Coarse) & BSD500(Coarse)  \\ \hline
		GSP       & 0.1237          & 0.1149          & 0.3225                & 0.3087                & 0.3067              & 0.3554          \\ \hline
		K-MEANS   & 0.1293          & 0.1252          & 0.3044                & 0.3105                & 0.3124              & 0.3154          \\ \hline
		MLG-SVD & {\bf 0.1326}    & {\bf 0.1366} & {\bf 0.3344}          & {\bf 0.3394}          & {\bf 0.3335}        & {\bf 0.3743} \\ \hline
		MLG-CP    & 0.1321          & 0.1293          & 0.3195                & 0.3256                & 0.3243              & 0.3641          \\ \hline
		IIC       & \multicolumn{5}{l|}{\multirow{4}{*}{}}                                                                  & 0.2071          \\ \cline{1-1} \cline{7-7} 
		GS        & \multicolumn{5}{l|}{}                                                                                   & 0.3658          \\ \cline{1-1} \cline{7-7} 
		BP        & \multicolumn{5}{l|}{}                                                                                   & 0.3239          \\ \cline{1-1} \cline{7-7} 
		DFC       & \multicolumn{5}{l|}{}                                                                                   & {0.3739}    \\ \hline
	\end{tabular}
	\label{BSD11}
\end{table*}

\begin{algorithm}[t]
	\begin{algorithmic}[1] 
		\caption{MLG-based Unsupervised Image Segmentation}\label{basic4}
		\STATE {\bf{Input}}: RGB Image $\mathbf{I}\in\mathbb{R}^{P\times Q\times 3}$;
		\STATE Build $N$ superpixels for the image $\mathbf{I}$ and calculate the value of superpixel based on the mean of all pixels inside that superpixel, i.e., $\mathbf{s}\in\mathbb{R}^{N\times 3}$;
		\STATE Construct a multilayer graph $\mathbf{A}\in\mathbb{R}^{M\times N \times M \times N}$;
		\STATE Find entity-wise spectrum $\mathbf{E}=[\mathbf{e}_1,\cdots,\mathbf{e}_N]\in\mathbb{R}^{N\times N}$;
		\STATE Select the first $K$ important leading spectrum based on the eigenvalues (singular values) of $\mathbf{E}$ as $\mathbf{C}\in\mathbb{R}^{N\times K}$;
		\STATE Cluster each row of $\mathbf{C}$, and assign the $i$th superpixel into $j$th cluster if the $i$th row of $\mathbf{C}$ is clustered into $j$th group;
		\STATE Assign all pixels inside one superpixel to the cluster of that superpixel;
		\STATE  {\bf{Output}}: The segmented image.
	\end{algorithmic}
\end{algorithm}

To test the proposed Algorithm 1, we first compare 
its results with those from
GSP-based method and traditional $k$-means clustering by
using a public BCCD blood cell dataset shown in Fig. \ref{CELL11}.
In this dataset, 
there are mainly three types of objects, i.e., White Blood Cell (WBC) vs. Red Blood Cell (RBC) vs. Platelet (P). 
We set the number of clusters to 3 and $N=1000$. For 
GSP-based spectral clustering, we construct graphs based on the Gaussian model by using information from all 3 color values 
$\sum_{\ell=1}^3|\mathbf{s}_i(\ell)-\mathbf{s}_j(\ell)|^2$ to
form edge connections in a single graph. There is only a single
$\delta_\ell$ and $t_\ell$ in the Gaussian model. 
For M-GSP, we use the MLG singular vectors (MLG-SVD), and tensor factorization (MLG-FZ) for spectral clustering, separately. 
Their respective results are compared in Fig \ref{CELL123}.
{WBCs are marked yellow, and RBCs are marked green. Platelet (P) is marked blue.}
From the 
illustrative results, 
MLG methods 
exhibit better robustness and are better in detecting regions under
noise.
Comparing results from different MLG-based methods, we find MLG-FZ to
be less stable than HOSVD, partly due to approximation algorithms 
used for tensor factorization. Overall, MLG-based methods shows 
reliable and robust performance over GSP-based method and $k$-means.

In addition to visual inspection of results for such images, we are 
further interested to numerically evaluate the performance of the proposed methods against some state-of-art methods for several more complex datasets
that contain more classes. For this purpose, we test our methods on the BSD300 and BSD500 datasets \cite{d13}.
{BSD300 contains 300 images with
labels, and BSD500 contains 500 images with labels.}
We first cluster each image, and label each cluster with the best map of cluster orders against
the ground truth. Numerically,
we use mIOU (mean of Intersection-over-Union), also known as the mean Jaccard Distance, for all clusters in each image to measure the performance. The Jaccard Distance between two groups $A$ and $B$ is defined as
\begin{equation}
	J(A,B)=\frac{|A\cap B|}{|A\cup B|}.
\end{equation}
A larger mIOU indicates stronger performance. To better illustrate the results, we considered two setups of datasets, i.e., one containing
fewer classes (coarse) and one containing all images (all). 
We compare our methods together with
$k$-means, GSP-based spectral clustering, invariant information clustering
(IIC) \cite{b1}, graph-based segmentation (GS) \cite{b2}, back propagation (BP) \cite{b3} and differentiable feature clustering (DFC) \cite{b4}.
The best performance is marked in bold.
From the results of Table \ref{BSD11}, we can see that
larger number of clusters in the first two columns generate worse performance.
There are two natural reasons. First, the mapping of the best order of cluster labels is more difficult for more classes. Second, the graph-based spectral clustering is sensitive to the number of $K$ leading spectra and the structure of graphs. 
Regardless, MLG-based methods still 
demonstrate better performance. 
Moreover, even when we use the same total number of nodes in a single layer graph and multilayer graph for another fairness comparison in terms of complexity, 
i.e., $N=300$ for graph and $N=100$ for MLG, 
MLG-based methods still perform better than graph-based methods in this example application.
MLG methods have competitive performances 
compared to the state-of-the-art methods. 
Note that, under proper training,
neural network (NN)-based methods may give
good results in cases with many clusters as suggested in \cite{b4}.

\subsection{Semi-Supervised Classification}

Semi-supervised classification is an important practical application
for {IoT intelligence}. 
In this application, we apply MLG polynomial filters 
for semi-supervised classification.
Traditional GSP defines adaptive filter
as 
\begin{equation}
    f(\mathbf{s})=\sum_i a_i\mathbf{W}^i\mathbf{s},
\end{equation}
where $\mathbf{W}$ is an adjacency matrix based on 
pairwise distance or a 
representing matrix constructed from the adjacency matrix.
Here,  signals are
defined as labels or confidence values of nodes, i.e., $\mathbf{s}=[\mathbf{s}_{L}^{\mathrm{T}} \: \mathbf{0}_{UL}^{\mathrm{T}}]^{\mathrm{T}}$
by setting unlabeled signals to zero. 
To estimate parameters $a_i$ of $f(\cdot)$,
Optimization can be formulated to minimize,
e.g., 
the mean square error (MSE) from ground truth label
$\mathbf{y}_L$
\begin{equation}
\min_\mathbf{a}\quad ||M(f(\mathbf{s}))_L-\mathbf{y}_L||_2^2,
\end{equation}
where $M(\cdot)$ is a mapping of
filtered signals to their discrete labels. For example, in a $\{\pm 1\}$ binary classification, one can assign a label to a 
filtered signal against a threshold (e.g. $0$).
Some other objective functions include 
labeling 
uncertainty, Laplacian regularization,
and total variation. Using estimated parameters, 
we can filter the signal one more time
to determine
labels for some unlabeled data by following the same process.

\begin{figure}[t]
	\centering
	\includegraphics[width=2in]{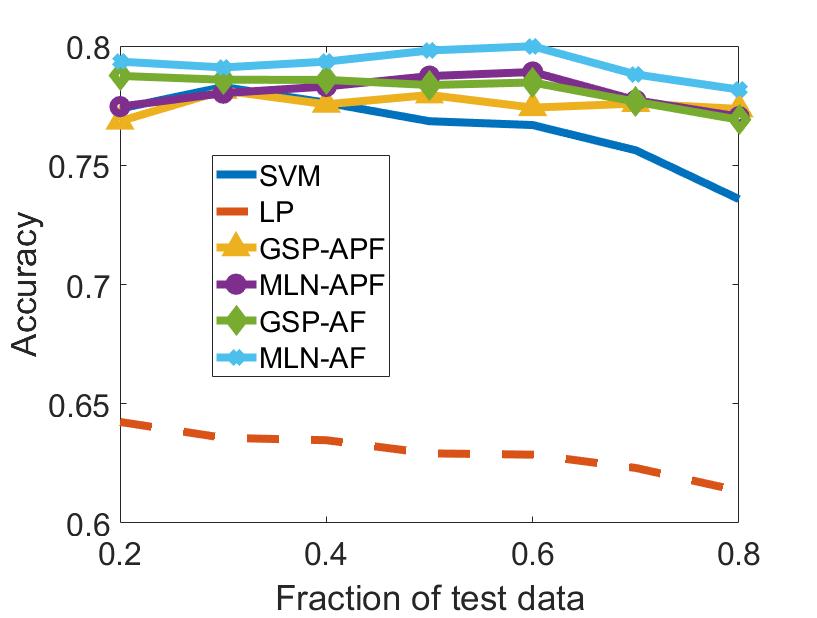}
	\caption{Results of Classification.}
	\label{cl}
\end{figure}

Similarly, in an MLG, we can also apply polynomial filters for label estimation. Given an arbitrary dataset $\mathbf{X}=[\mathbf{x}_1,\cdots,\mathbf{x}_N]\in\mathbb{R}^{K
	\times N}$ with $N$ signal points and $K$ features for each node, we can construct a MLG by 
defining $M=K$ layers based on features and $N$ entities based on signal points.
The inter- and intra- layer connections are calculated by the Gaussian distance with different parameters.
Let its
adjacency tensor
$\mathbf{A}\in\mathbb{R}^{M\times N\times M \times N}$. A signal is defined by
\begin{equation}\label{sss}
\mathbf{s}=\begin{bmatrix}
\mathbf{s}_{L}&\cdots&	\mathbf{s}_{L}\\
\mathbf{0}_{UL}&\cdots&\mathbf{0}_{UL}
\end{bmatrix}^{\mathrm{T}}\in\mathbb{R}^{M\times N},
\end{equation}
which is an extended version of graph signal. Note that we do not necessarily need to order signals by placing zeros in the rear. 
We only write the signal as Eq. (\ref{sss}) for notational
convenience. We now apply polynomial filters on signals, i.e.,
\begin{equation}
	\mathbf{s}_1=h_1(\mathbf{s})=\sum_i a_i\mathbf{A}^{[i]}\diamond\mathbf{s},
\end{equation}
and
\begin{equation}
	\mathbf{s}_2=h_2(\mathbf{s})=\mathbf{A}^{[i]}\diamond\mathbf{s}.
\end{equation}
For a filtered signal $\mathbf{s}_X\in\mathbb{R}^{M\times N}$ {($X=1,2$)}, 
we define a function to map 2-D signals into 1-D
by calculated the column-wise mean of $\mathbf{s}_X$, i.e.,
\begin{equation}
	\mathbf{\bar s}_X={\rm mean}_{col}(\mathbf{s}_X)\in\mathbb{R}^{1\times N}.
\end{equation}
\begin{figure}[t]
	\centering
	\includegraphics[width=3in]{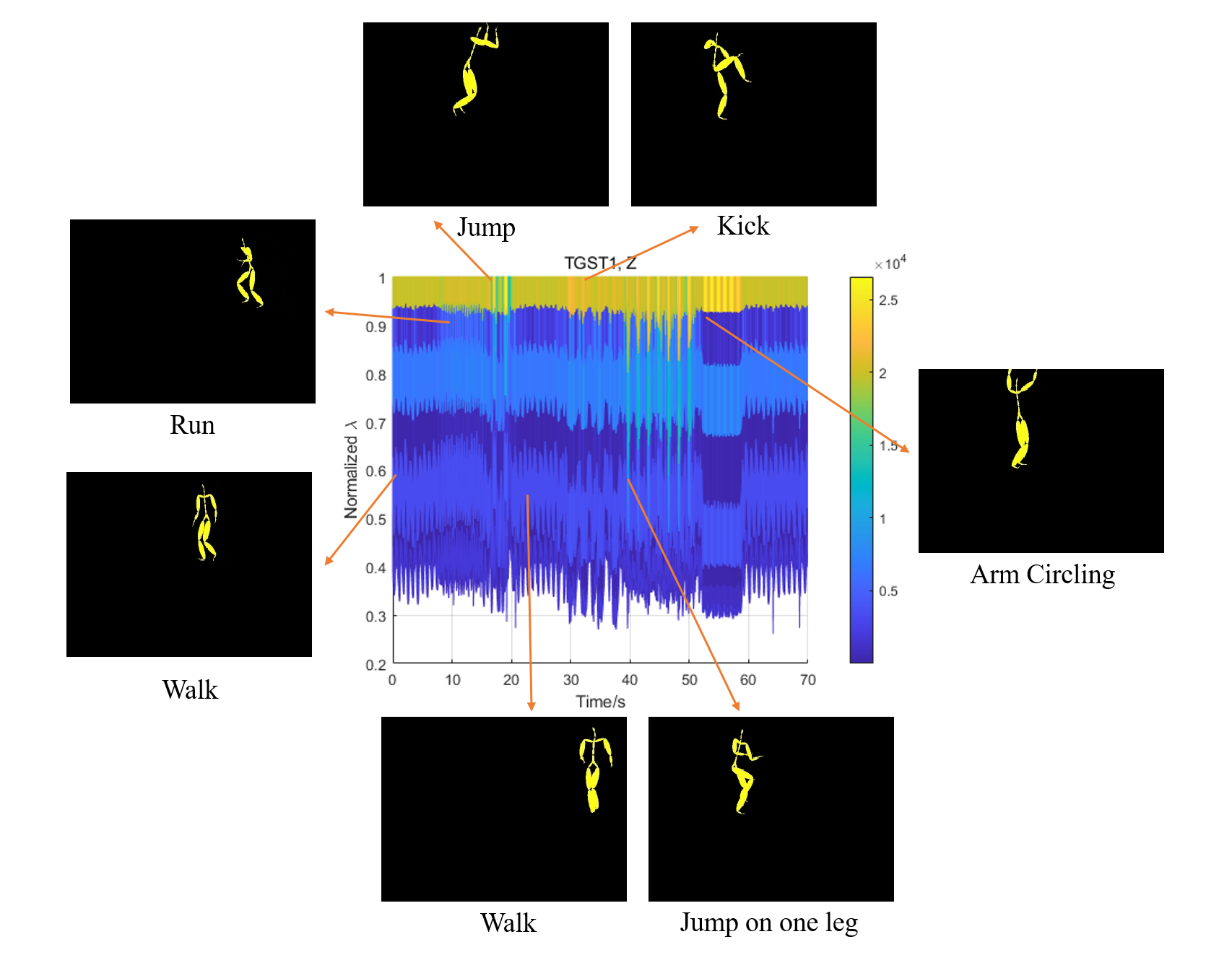}
	\caption{Example of Transformed Signals in a Dynamic Point Cloud.}
	\label{ex}
\end{figure}
Next, we can define a function $M(\cdot)$ on $\mathbf{\bar s}_X$ and 
consider certain objective functions in
filter design. 
To validate the efficacy of polynomial filtering in the MLG framework, 
we test $h_1(\cdot)$ and $h_2(\cdot)$ for the binary classification 
problem on the Cleveland Heart Disease Dataset. 
In this dataset, there are $297$ data points with 13 feature dimensions.
We directly build a MLG with $N=297$ nodes in each of
the $M=13$ layers. More specifically, 
we directly use the labels as $\mathbf{s}$.
For $h_1(\cdot)$ (AF), we set $a_i\neq 0$ for at least one $i>0$.
Using MSE as objective function,
we apply a greedy algorithm to estimate parameters $\{a_i\}$. 
We limit the highest polynomial order to ${10}$.
For $h_2(\cdot)$ (APF), we estimate a classification
threshold via the mean of $\bar{ \mathbf{s}}_X$ 
by setting the polynomial order $i=10$.

We compare our methods with GSP-based method in similar setups as in aforementioned examples. The only difference is that we use $\bar{ \mathbf{s}}_X$ in M-GSP and use
$\mathbf{s}'=f([
\mathbf{s}_{L}^T\; 
\mathbf{0}_{\rm UL}^T ]^T)$
in GSP for mapping and classification. We also present the results of label propagation and SVM for comparison. We randomly split the test and training data for 100 rounds. From the results shown in Fig. \ref{cl}, 
GSP-based and M-GSP based methods exhibit
better performance than traditional learning algorithms, 
particularly when the fraction of test samples
is large. In general, M-GSP based methods demonstrate
superior performance 
among all methods owing to 
its strength to extract `multilayer' features, {which could  potentially benefit semi-supervised classification tasks in IoT systems.}

\subsection{Dynamic Point Cloud Analysis}

\begin{figure}[t]
	\centering
	\includegraphics[width=3in]{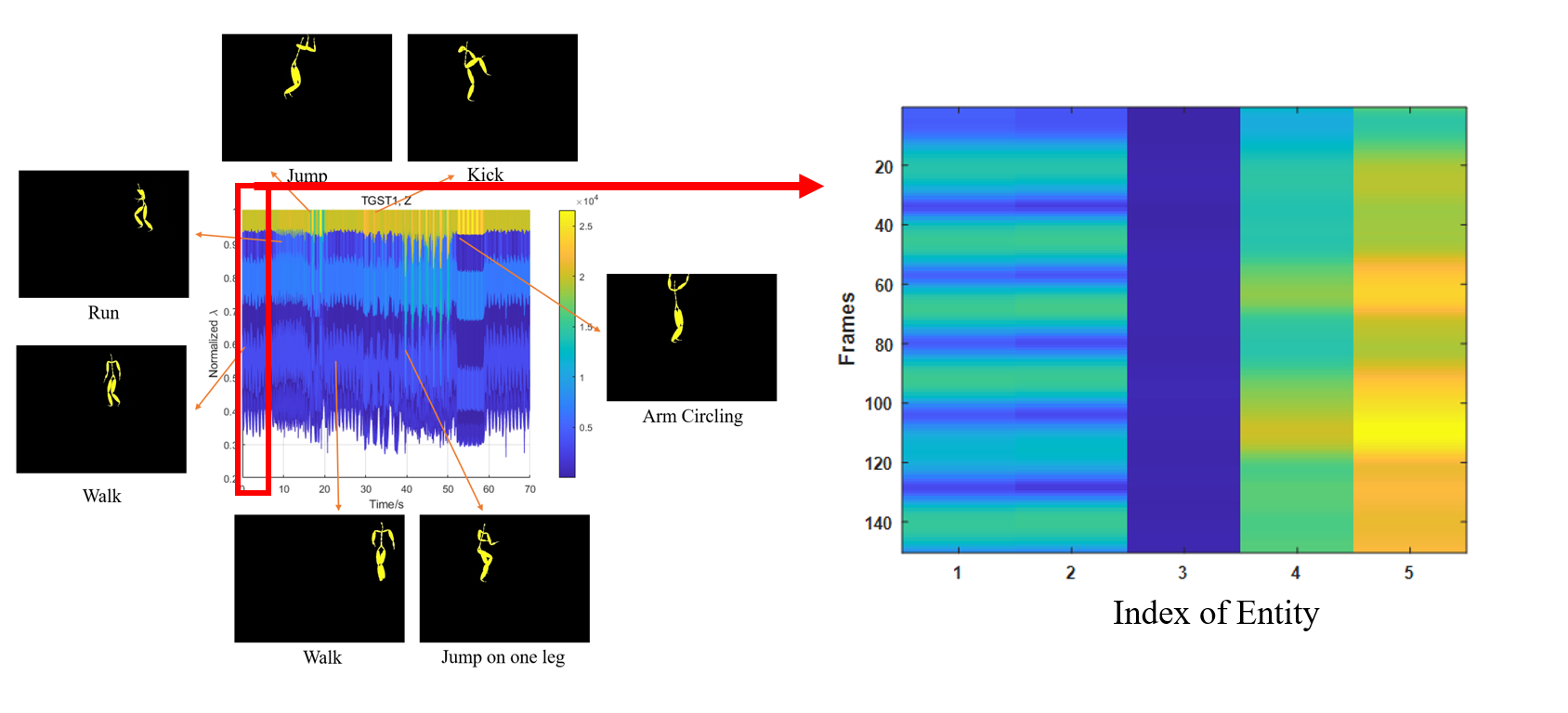}
	\caption{Example of Filtered Signals in a Dynamic Point Cloud: entity 1 and entity 2 are legs; entity 3 is head; entity 4 and entity 5 are hands.}
	\label{ex111}
\end{figure}

{3D perception plays an important role in the high
	growth fields of IoT devices and cyber-physical systems,
	and continues to drive many progresses made in advanced
	point cloud processing \cite{q1}.}
Here, we propose a short time M-GST method to analyze 
dynamic point cloud. Given a dynamic point cloud with $M$ frames and at most $N$ points in each frame, we model it as a multilayer graph with $M$ layers and $N$ nodes in each layer.
More specifically, we test the singular spectrum analysis over the motion sequences of subject 86 in the CMU database \cite{CMU}. To implement the M-GST, we first divide
the motion sequence into several shorter sequences with $N_f$ frames. Next
for each shorter sequence, we model interlayer connections by 
connecting points with the same label among
successive frames. For points in the same frame, we connect 
two points based on the Gaussian-kernel within
a Euclidean threshold $\tau_s$ \cite{c6}. 
Let $\mathbf{x}_i$ be the 3D coordinates of the $i$th point. We
assign an edge weight between two 
points $\mathbf{x}_i$ and $\mathbf{x}_j$ as a nonzero
$A_{ij} = \exp(-\Vert \mathbf{x}_i-\mathbf{x}_j \Vert^2_2/{\sigma^2}
)$ only if $\Vert \mathbf{x}_i-\mathbf{x}_j \Vert^2_2\leq\tau_s$.
Next, we estimate the spatial and temporal basis vectors of each shorter-term sequences by HOSVD in Eq. (\ref{decomposeS}). Finally, we use the 3D coordinates of all points in each shorter-term sequences as signals and calculate their M-GST.
To illustrate the results of M-GST, we examine
the spectrogram similar to that of short-time Fourier transform (STFT) \cite{c36}.
In Fig. \ref{ex}, we transform the signal defined by the coordinates in $Z$ dimension via M-GST and illustrate the transformation results for the divided frame sequence. From Fig. \ref{ex}, one can easily identify different motions based on the MLG singular analysis. 

{To explore motions in dynamic point clouds, we can also apply the entity-wise MLG highpass filters described in {Section \ref{fefi}} to capture critical details of human bodies. More specifically, we select the first 140 frames in `walking' and define the norm of three coordinates as signals. We select 5 body joints (entities) in each temporal frame (layers) shown as Fig. \ref{ex111}. From the results shown, entity 1 and entity 2 exhibit periodic patterns which are linked to the leg motion. Entity 3 (head) shows little movement relative to the main body. Entity 4 and entity 5 (hands) display more irregular patterns since they do not directly identify `walking'. To summarize, the MLG highpass filter can efficiently capture some key information of body movement and identify the meaning of nodes (entities). These and related information can assist further analysis of dynamic point clouds including compression and classification.}

{Our future works shall target more practical applications of point cloud on IoT devices, including point cloud compression, low-complex point cloud segmentation and robust denoising.}

\subsection{Other Potential Applications in IoT Systems}
{
Along with the widespread deployment
of IoT technologies, system
structures become increasingly complex. 
Traditional graph-based tools are less
adept at modeling 
`multilayer' graph interactions. The
more general model of M-GSP provides additional opportunities for IoT applications. Here, we suggest several potential scenarios in IoT systems for M-GSP:
\begin{itemize}
	\item IoT networks with multilayer structure fit naturally to MLG, 
	for which M-GSP can be designed for various tasks such as intrusion detection, resource management and state prediction;
	\item Because of the dynamic
	nature in practical IoT networking, even signals on single-layer IoT systems naturally fit a spatial-temporal graph model, which can be also characterized by MLG. For such dynamic IoT systems, M-GSP tools, such as adaptive filters and MLG learning machines, can be developed for signal prediction and node classification.
\end{itemize}
Overall, the power of MLG in extracting underlying `multilayer/multi-level' structures in the IoT systems makes M-GSP a potentially important tool in handling high-dimensional signal processing and learning tasks.}

\section{Conclusion} \label{con}
In this work, we present a novel tensor-based
framework of multilayer graph signal processing (M-GSP) that 
naturally generalizes the traditional GSP to multilayer graphs. We first
present the basic foundation and definitions of M-GSP including MLG signals, signal shifting, spectrum space, singular space, and filter design.
We also provide interpretable discussion 
and physical insights through numerical results and examples to 
illustrate the strengths, general insights, 
and benefits of novel M-GSP framework. 
We further demonstrate exciting potentials of
M-GSP in data processing
applications through
experimental results in several practical scenarios.

{With recent
advances in tensor algebra
and multilayer graph theory, more opportunities are emerging to
explore M-GSP and its applications. One such interesting problem is the
efficient construction of multilayer graph, where M-GSP spectrum properties could improve the robustness of graph structure. Another promising direction is to develop multilayer graph neural networks based on the M-GSP spectral convolution. Additional future directions include the development of M-GSP sampling theory and fast M-GFT.}

\appendix
Unlike for undirected graphs, representing tensors of 
directed graphs is asymmetric, thereby making each layer or entity characterized by a pair of spectral vectors. To find the spectrum
space of a directed multilayer graph, we also 
present two ways to compute: 
\begin{itemize}
	\item \textit{Flattening analysis:}   Similar to the representing tensor of  undirected graphs, we flatten the representing tensor as a second-order supra-matrix, and define spectrum space as the {reshaped} eigenvectors of the supra-matrix. The flattened matrix $\mathbf{A}_{FX}$ (or $\mathbf{A}_{FN}$, $\mathbf{A}_{FL}$) can be decomposed as 
	\begin{equation}
	\mathbf{A}_{FX}\approx\mathbf{E}\Sigma \mathbf{E}^{-1},
	\end{equation}
	where $\mathbf{E}\in\mathbb{R}^{MN\times MN}$ is the matrix of 
	eigenvectors and $\Sigma=diag(\lambda_i)$ is a diagonal matrix of eigenvalues. Then, we can reshape the eigenvectors, i.e., each column of $\mathbf{E}$ as $\mathbf{V}_k\in\mathbb{R}^{M\times N}$, and reshape each row of $\mathbf{E}^{-1}$ as $\mathbf{U}_k\in\mathbb{R}^{M\times N}$.
	Consequently, the original tensor can {be} decomposed into
	\begin{equation}
	\mathbf{A}\approx\sum_{k=1}^{MN}\lambda_k\mathbf{V}_k\circ \mathbf{U}_k.
	\end{equation}
	\item \textit{Tensor Factorization}: We can also compute the spectrum from the tensor factorization based on CP-decomposition
	\begin{align}\label{cp1}
	\mathbf{A}&\approx\sum_{k=1}^{R}\lambda_k\mathbf{a}_k\circ\mathbf{b}_k\circ\mathbf{c}_k\circ\mathbf{d}_k\\
	&=\sum_{k=1}^{R}\lambda_k\mathbf{V}_k\circ \mathbf{U}_k.
	\end{align}
	where $R$ is the rank of tensor, $\mathbf{a}_k,\mathbf{c}_k\in\mathbb{R}^{M}$ characterize layers, $\mathbf{b}_k,\mathbf{d}_k\in\mathbb{R}^{N}$ characterize entities, and $\mathbf{V}_k=\mathbf{a}_k\circ\mathbf{b}_k, \mathbf{U}_k=\mathbf{c}_k\circ\mathbf{d}_k\in\mathbb{R}^{M\times N}$ characterize the joint features. 
	Since there are $MN$ nodes, it is clear that $R\leq MN$. Note that, 
	for a single layer, Eq. (\ref{cp1}) reduces to 
	\begin{align}\label{cp2}
	\mathbf{A}\approx\sum_{k=1}^{N}\lambda_k\mathbf{v}_k\circ \mathbf{u}_k.
	\end{align}
	Moreover, if $\mathbf{V}=(\mathbf{v}_k)$ and $\mathbf{U}=(\mathbf{u}_k^\mathrm{T})=\mathbf{V}^{-1}$ are orthogonal bases, Eq. (\ref{cp2}) is in a consistent form of the {eigen-decomposition} in a single-layer normal graph. In addition, Eq. (\ref{cp_dec}) is also a special case of Eq. (\ref{cp1}) if the multilayer graph is undirected.
\end{itemize}

Since tensor decomposition is less stable when exploring the factorization of 
a specific order or when
extracting the separate features in the asymmetric tensors, we will defer more general
analysis of directed networks to future works.

\vfill

\end{document}